Spring 5-10-2019

# Maine's Forestry and Logging Industry: Building a Model for Forecasting

Jonathan Gendron
*University of Maine*, jonathan.e.gendron@maine.edu

Follow this and additional works at: https://digitalcommons.library.umaine.edu/etd

### Recommended Citation

Gendron, Jonathan, "Maine's Forestry and Logging Industry: Building a Model for Forecasting" (2019). *Electronic Theses and Dissertations*. 2998.
https://digitalcommons.library.umaine.edu/etd/2998

**MAINE'S FORESTRY AND LOGGING INDUSTRY:**

**BUILDING A MODEL FOR FORECASTING**

By

Jonathan E. Gendron

B.S., Towson University, 2016

A THESIS

Submitted in Partial Fulfillment of the

Requirements for the Degree of

Master of Science

(in Economics)

The Graduate School

The University of Maine

May 2019

Advisory Committee:

    Andrew J. Crawley, Assistant Professor of Regional Economic Development, Advisor

    Adam Daigneault, Assistant Professor of Forest, Conservation, and Recreation Policy

    Kathleen P. Bell, Professor of Economics

# MAINE'S FORESTRY AND LOGGING INDUSTRY:

# BUILDING A MODEL FOR FORECASTING

By Jonathan E. Gendron

Thesis Advisor: Dr. Andrew J. Crawley

An Abstract of the Thesis Presented
in Partial Fulfillment of the Requirements for the
Degree of Master of Science
(in Economics)
May 2019

From 2000 to 2017, 64% of Maine's pulp and paper processing mills shut down; these closures resulted in harmful effects to communities in Maine and beyond. One question this research asks is how will key macroeconomic and related variables for Maine's forestry and logging industry change in the future? To answer this, we forecast key macroeconomic and related variables with a vector error correction (VEC) model to assess past and predict future economic contributions from Maine's forestry and logging industry. The forecasting results imply that although the contribution of the industry in Maine would likely remain stable due to level prices and a slight increase in output, local Maine communities could be worse off due to decreases in employment and firms. We then incorporated these forecasts into a 3-stage modeling process to analyze how a negative shock to exchange rates from an increase in tariffs could affect Maine's employment and output. Our results suggest that increased tariffs will reduce output and increase employment volatility in Maine. Rising uncertainty and costs of business operations suggest care should be taken when changing tariffs and trade restrictions, especially when changes to business operations can harm markets and communities.

# ACKNOWLEDGEMENTS

I sincerely thank everyone who has helped me through the journey of graduate studies. Thank you so much to the members of my committee for generously providing their time, expertise, and encouragement. Thank you, Adam, for taking this economist into the forestry industry with kindness, enthusiasm and open arms. Thank you, Andy and Kathleen, for not only helping me academically, but also for your invaluable personal guidance and help through this journey. And Andy, thank you for regularly challenging me to do my best, and providing so much positive encouragement. I wish to also express my thanks for all the support from faculty and fellow colleagues in the School of Economics. I cannot think of anywhere else where everyone is so kind, intelligent, selfless, and welcoming. Of course, thank you to my friends and family. Lastly, a special thanks to my family for your love and encouragement, that encouragement lead me on this fantastic journey at the University of Maine.

# TABLE OF CONTENTS

# LIST OF TABLES

# LIST OF FIGURES

# LIST OF ACRONYMS

US: United States

NAICS 113: Industry classification of the forestry and logging industry (logs, not lumber)

NAICS 321: Wood products processed in a mill (such as lumber and furniture)

GDP: Gross Domestic Product

RGDP: Real Gross Domestic Product

LQ: Location Quotient

IRF: Orthogonalized Impulse Response Function

CVDAD: Countervailing Duty and Anti-Dumping Duty

VEC: Vector Error Correction [Model]

VAR: Vector Auto Regressive [Model]

AR: Auto Regressive [Model]

MA: Moving Average [Model]

ARIMA: Autoregressive Integrated Moving Average [Model]

2SLS: Two Stage Least Squares [Model]

3SLS: Three Stage Least Squares [Model]

CGE: Computable General Equilibrium

ADF: Augmented Dickey-Fuller

# CHAPTER 1: INTRODUCTION

## Background

The forestry and logging industry includes the processes of growing trees, harvesting logs, transporting logs directly to buyers or processing mills (that convert logs to an end product), and the support staff throughout these stages of harvest and transport (Forest Opportunity Roadmap Maine, 2018). It is important to mention that logs are unprocessed (before arriving to a mill), lumber is processed at mills, and timber is a broad term that includes logs and timber (USDA, 2009). Like any industry, the forestry and logging industry creates jobs and supplies demanded goods and services internationally, nationally, regionally, and locally. In 2017, forestry, fishing, and related activities (which includes the forestry and logging sector) accounted for 596,000 United States (US) employees, $36.4 billion value added in the US, and $54.7 billion of gross output for the US (BEA, 2017). This includes when trees are planted to when they are harvested, and the logs are on their way to a mill for processing. In contrast, wood products accounted for $112.2 billion of US gross output, 398,000 US employees, and $38.1 billion value added in the US (BEA, 2017). This includes once logs arrive at a mill and are processed into lumber or wood products. We will use the term forest products industry to describe the sector that processes logs into lumber or another end product in a mill. In addition, softwood lumber trade between the US and Canada annually makes up $4 to $7 billion of goods. Indirect multiplier effects also have a strong impact. For instance, the initial investment and employment of a new mill are a positive direct effect that also indirectly increases the town's tax base because of new employees with new wages that are spent on local goods and services (Crandall et al., 2017). These indirect effects further increase the positive (or negative) effect to the local economy.

In the past 20 years both the forest product industry and the forestry and logging industry in the US have experienced numerous challenges due to the rise of electronic media, the rise of competing suppliers, and big changes in the manufacturing industry (Irland, 2017). Decreases in lumber demand, mill shutdowns, and trade disputes with Canada complicate business operations. Automation that took off in the 1990s decreased demand for manufacturing employment and globalization of manufacturing decreased US manufacturing output (Irland, 2017). During this time, the rise of electronic media in the 1990s (including e-mail, e-publishing and social media) lowered demand for print media (such as letters, newspaper, etc.). Newsprint fared the worst declining 55% between 2000-2014 in North America (Irland, 2017). These factors led to many plant closures and consolidations. In the state of Maine, 64% (11 out of 17) of the pulp and paper mills shut down since 2000 (Crandall et al., 2017). These shutdowns happened throughout Maine at different time periods (see Figure 1).

Figure 1. Map of Pulp and Paper Mill Shutdowns in Maine. Mill shutdowns that occurred in Maine since 2000.

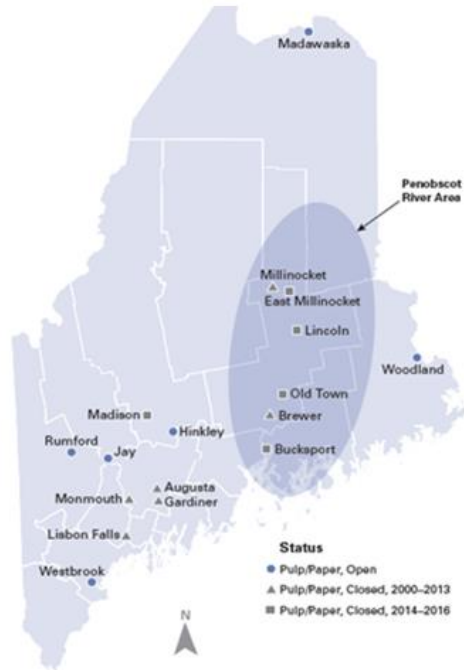

These shutdowns have negatively impacted rural mill towns, where the mills made up majority of the towns' tax bases. In 2001 there were 5 towns where 60% or more of the tax base was tied to their mill (Irland, 2001). Canada and the US have historically had trade disputes since the 1980s', those disputes have resulted in a range of tariffs from 10% to 28%. From August 2001 to September 2006 the US implemented Countervailing Duty and Anti-Dumping Duty (CVDAD) tariffs on Canada, which lead to the largest reduction in US imports of 13% (Zhang & Parajuli, 2016).

    The US government has recently been enacting new tariffs and trade restrictions for various industries, and it is very important to understand the impacts this will have. Analysis of the most recent tariffs and trade restrictions on steel and aluminum (25% and 10% respectively) enacted by the US government will help show the negative effects of these new tariffs and trade

restrictions (even though they are primarily unrelated to forestry and logging). Although these steel and aluminum tariffs and trade restrictions were enacted to create US jobs, these instead caused a net job loss nationwide. For each job created 16 jobs were lost, which totaled 402,445 jobs lost nationwide (Francois, Baughman, & Anthony, 2018). President Clintons' economic advisor Dr. Laura Tyson voiced her concern that tariffs and other trade restrictions could easily lead to potential retaliation tariffs in response, higher input prices, and global trade slowing down (Tyson, 2019). This effects all industries, Francois et al. (2018) estimate annual gross domestic product (GDP) will decrease by $36.8 billion in response to this steel and aluminum tariff and trade restriction change. These chapters aim to best understand the direction of key macroeconomic and related variables, as well as impacts to the forestry and logging industry if there are changes to tariffs and other trade restrictions.

**Motivation**

The forestry and logging industry significantly impact international, national and local economies. Historically, this industry has been especially important to Maine, where many local economies and communities have heavily relied on this industry for their livelihood. In the last 20 years, both the forestry and logging industry and the forest products industry have been devastated by the negative impacts that occurred due to the rise of electronic media, the rise of competing suppliers, and big changes in manufacturing in general. Negative impacts include decreases in lumber demand, mill shutdowns, and trade disputes with Canada. Recently, the US has been renegotiating trade restrictions with various countries throughout the world. Both the forestry and logging industry as well as the forest products industry have been adapting to the associated historic trade shocks in various ways. My thesis research addresses two questions: (1) How will key macroeconomic and related variables for Maine's forestry and logging industry change in the future? (2) How will shocks to exchange rates from an increase in tariffs (and other trade restrictions) affect employment and output in Maine's forestry and logging industry?

**Thesis Organization**

Chapters 2 and 3 present two related analyses of Maine's forestry and logging industry (and the closely related forest products industry); both analyses use a macroeconomic forecasting and shock analysis approach. Chapter 2 forecasts key macroeconomic and related variables for the forestry and logging industry in Maine and attempts to answer question 1. To answer question 2, Chapter 3 makes use of the forecasting economic model developed in Chapter 2 examines how an exchange rate shock would impact employment and output in the forestry and logging industry in Maine.

# CHAPTER 2: MAINE'S FORESTRY AND LOGGING INDUSTRY: FORECASTING WITH A VEC MODEL

## Introduction

The forestry and logging industry contributes to national and local economies through job creation and the supply of demanded goods and services. Jobs are created in this industry at multiple stages throughout the processes of growing trees, harvesting logs, and transporting logs. The forestry and logging industry includes the processes of growing trees, harvesting logs, transporting logs directly to buyers or processing mills (that convert logs to an end product), and the support staff throughout these stages of harvest and transport (Forest Opportunity Roadmap Maine, 2018). We will also use the term forest products industry to describe the sector that processes logs into lumber or other end products in a mill. Many towns throughout the state of Maine have suffered when their mills shut down, but there are also towns with mills still in operation that continue to depend on employment associated with mills, as well as harvesting and logging (Weeks, 1990). In the past 20 years, 64% (11 out of 17) of the pulp and paper mills across the state have closed, which has hurt many towns in Maine (Crandall et al., 2017). During 2001, five towns had mills that made up more than 60% of the local tax base (Irland, 2001). Mill shutdowns and tax base dependence can have negative economic impacts. The objective of this paper is to address these concerns explicitly for the forestry and logging industry in the state of Maine. To do so, this research builds a model with the capacity to forecast key macroeconomic and related variables in Maine's forestry and logging industry.

## The Forestry and Logging Industry in the US

In 2017, forestry, fishing, and related activities accounted for 596,000 US employees, $36.4 billion value added in the US, and $54.7 billion of gross output for the US (BEA, 2017).

This includes the forestry and logging industry as well as fishing and other activities grouped together in NAICS 11. In contrast, wood products accounted for $112.2 billion of US gross output, 398,000 US employees, and $38.1 billion value added in the US (BEA, 2017). This includes once timber arrives at a mill and is processed into lumber or wood products. Since the mid-1990s, the rise of electronics (including e-mail, e-publishing and social media) has lowered demand for print media (such as letters, newsprint, etc.) (Irland, 2017). Between 2000-2014, North American newsprint declined the most, falling by 55% during this time period. Even though the demand for print media has fallen, demand for packaging has been increasing with shipping services (Berg & Lingqvist, 2017). In 2013, the majority of US paper companies experienced poor financial returns (Howard & Westby, 2013). In summary, the 1990s was a decade of low profitability (since the cost of capital exceeded revenue), and then the 2000s were a decade of mill closures and mill consolidations (Irland, 2017). Even critical investments to improve the cost efficiency of mills were not enough to help the decline of processing mills. Lumber is a critically important intermediate good in construction, as well as wood fuels, and furniture (U.S. Congress, 1983). So, when demand for these commodities change, the forestry and logging sector is greatly affected.

In 2016, the US consumed approximately 80.6 million cubic meters of softwood lumber (ForestEdge LLC & Wood Resources Int LLC, 2018). During 2013 the US was globally the largest exporter, second largest producer and consumer, and third largest importer of hardwood lumber (Luppold & Bumgardner, 2015). Canadian lumber suppliers have been a strong competitor of US suppliers, and general offshoring has caused US lumber output to decrease (Irland, 2017). US lumber output decreased further because of the collapse of housing construction in 2006 that resulted from the Great Recession (Woodall et al., 2011). In the 1990s,

US manufacturing employment decreased due to automation, and US manufacturing output decreased due to manufacturing globalization (Irland, 2017). Canada provides 94% of softwood lumber imports for the US, and the US imports about 48% of the amount produced by Canada (ForestEdge LLC & Wood Resources Int LLC, 2018). This shows that there is an important trade dependence between the US and Canada. According to Wear and Murray (2004) supply shocks from trade policies between Canada and the US reduced the supply of timber by about 15% during the early 2000s. Baek (2012) argued that since there is an ever-changing pattern of bilateral lumber trade between the US and Canada, it is critically important to best understand the effect of macroeconomic and market factors. Tariffs, which are essentially a trade restriction that changes the cost of buying or selling goods internationally, can have large impacts as well (York, 2018). US and Canadian softwood lumber trade annually makes up between $4 to $7 billion of goods, so even slight percentage changes in tariffs could have very large impacts to either or both countries (Zhang & Parajuli, 2016).

**The Forestry and Logging Industry in Maine**

In 1634 the first sawmill was built in Maine, and by 1830 Bangor Maine became the worlds' largest lumber shipping port (Judd et al., 2010). Over time, both the forestry and logging and forest products industries have moved to the Northwest and Southeast regions of the US. During this time, Maine also began to face competition from Canadian lumber suppliers, and automation led to a general decrease in demand for manufacturing employment (Irland, 2017). According to Smith et al. (2009), Maine is not only known for its' iconic forests, but also its' reliance on the forest products industry. In 2016, this $8.5 billion industry made up 5% of Maines' GDP and created 4.17% of Maines' jobs which employed 35,000 workers (Forest Opportunity Roadmap Maine, 2018). The same study found that this industry in Maine is

expected to increase 40% by 2025. Pulp products, paper products, and transportation equipment constitute the largest coverage of manufactured goods, and the manufacturing sector is the second largest piece of market value for Maine goods and services (Muskie et al., 2019). Maine has already faced numerous mill shutdowns in the past decade. Only 6 out of the 17 total mills are currently operating[1] (Crandall et al., 2017). These shutdowns have been so threatening to the economy of Maine that senators Susan Collins and Angus King labeled this situation as, "an economic crisis of unprecedented magnitude" (Fishell, 2016).

Paper mills that have not shut down have had to adapt to differences in demand due to the fall of print, the rise of electronic commerce, and the rise of competing suppliers (Berg & Lingqvist, 2017; Spelter, 2002). Maine specialized in production for printing and writing grades (which have faced decreased demand), and large modern mills were built in the Southeast and offshore (which have faced increased demand) (Irland, 2017). Even though general demand stabilized in the long run, these shocking transitions caused 64% of the mills in Maine to shut down (Crandall et al., 2017). For towns where the economy was focused on the mills (such as Millinocket), the businesses and people of these towns suffered (Weeks, 1990). In 2001, there were five towns where the mills accounted for 60% and more of the local tax base, one of those mills provided 85% of the towns' tax revenue (Irland, 2001). This goes to show that changes in both the forestry and logging industry as well as the forest products industry do not just have an important impact within the industry, but also the towns and people they serve.

---

[1] This does not include the mills in Old Town and Rumford Maine scheduled to reopen later this year.

**Forestry and Logging in Maine Compared to the US**

    To compare the importance of the forestry and logging industry in Maine compared to the US, we are going to use location quotients (LQ). Location quotients are a ratio of ratios which measure an industry in a region compared to that industry in the nation:

$$LQ = \frac{I_r/E_r}{I_n/E_n}$$

where $I_r$ is the industry in a region, $I_n$ is the industry in the nation, $E_r$ is employment in the region, and $E_n$ is employment in the nation. This is done by the finding the ratio of an industries regional employment share compared to an industries national employment share (BLS, 2011). If the LQ is greater than 1, then the industry is considered significant for the area. The forestry and logging industry in Maine has an LQ value ranging from 8 to 10, which means that this significant industry employed 8 to 10 times more people in Maine compared to the national rate (see Figure 2).

Figure 2. Graph of Maine's Forestry and Logging Industry LQ. Graph of LQ values during 2001-2017.

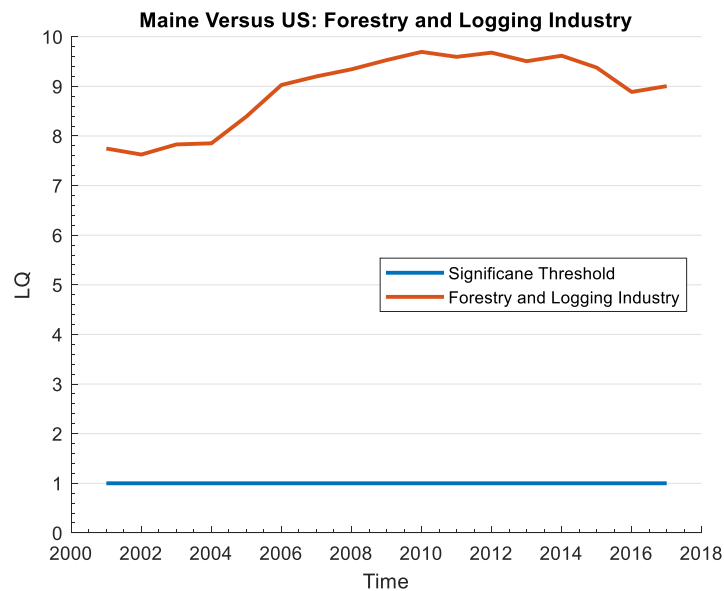

Figure 3. Graph of LQs for Maine's Other Industries. Graph of LQ values during 2001-2017.

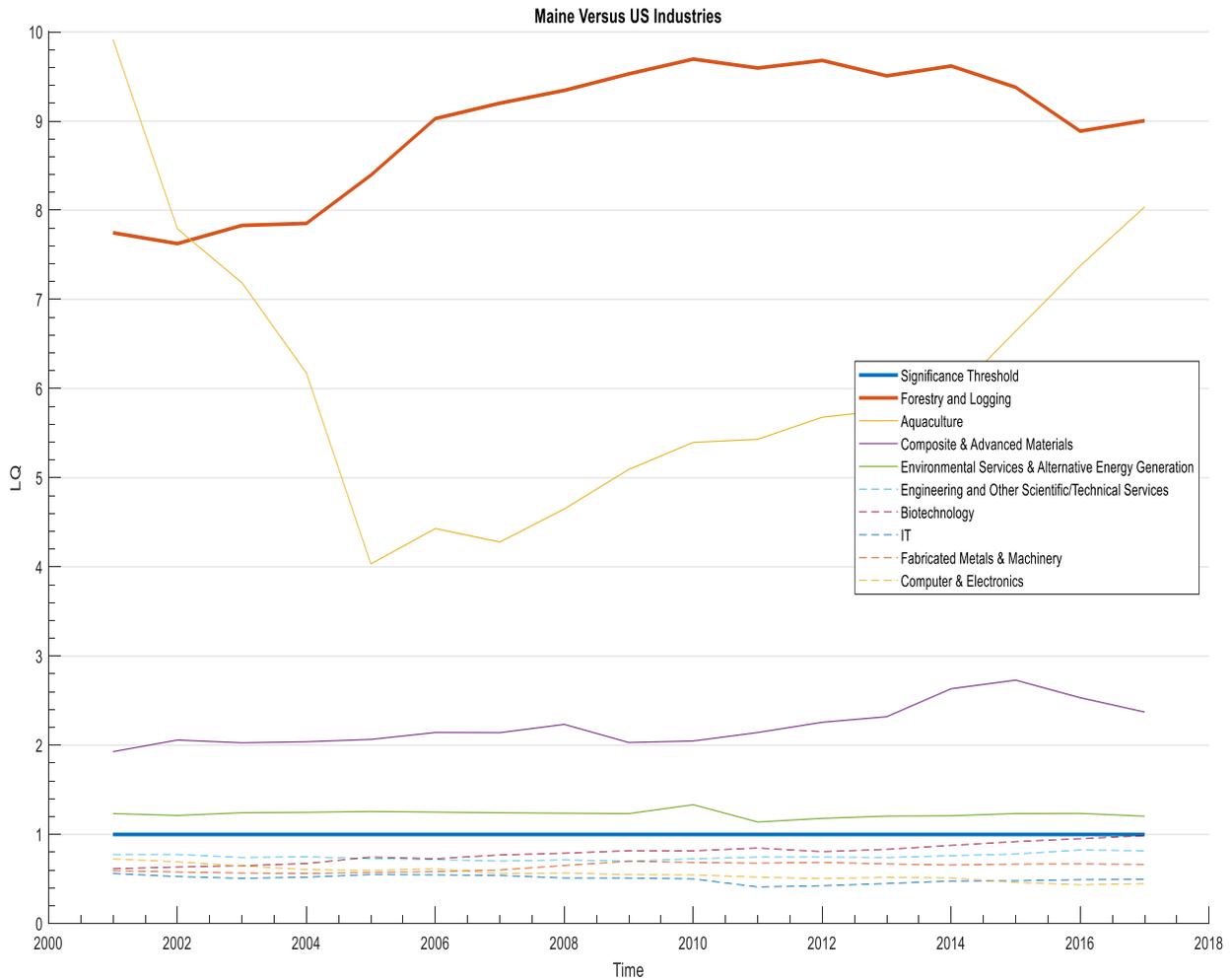

Figure 3 shows other important industries for Maine based on LQ values. This shows that environmental services/alternate energy generation, composite/advanced materials, and aquaculture are all more significant to Maine compared to the US. All the other industries in the graph are more significant in the US compared to Maine (since their LQs are less than 1). Composite/advanced materials forestry and logging industry and aquaculture are by far the strongest industries in Maine.

**Summary**

In 2017, forestry, fishing, and related activities accounted for 596,000 US employees, $36.4 billion value added, and $54.7 billion of gross output for the US (BEA, 2017). This includes when trees are planted as well as when they are harvested and the timber is on it's way to a mill for processing. In contrast, wood products accounted for $112.2 billion of US gross output, 398,000 US employees, and $38.1 billion value added (BEA, 2017). This includes once timber arrives at a mill and is processed into lumber or wood products. The US is globally one of the top consumers, importers and exporters of lumber (Luppold & Bumgardner, 2015).

Canada is a critical trading partner of the US, 94% of US softwood lumber imports are from Canada (which makes up 48% of Canada's production) (ForestEdge LLC & Wood Resources Int LLC, 2018). Background was also provided for the forestry and logging industry in Maine (and the related forest products industry). In 2016, Maine's forest products industry was valued at $8.5 billion (5% of Maine's GDP), employed 35,000 workers (4.17% of Maine's jobs), and this industry in Maine is expected to increase 40% by 2025 (Forest Opportunity Roadmap Maine, 2018). Since the 1990s, both the forestry and logging industry and forest products industry in both the US and Maine have changed due to general changes of manufacturing (as well as specific changes to supply and demand in the industries) (Irland, 2017).

**Literature Review**

Forecasting macroeconomic variables is a very common practice applied to many real-world issues. In fact, the Federal Reserve regularly uses such forecasts to make decisions that impact the US and global economy (Federal Reserve System, 2016). The most common forecasts are for employment, output, prices from inflation, and the federal funds rate. There are errors observed in any type of forecasting method, and forecast predictions are less accurate the further into the future they go. Regardless, forecasting estimates are analytic and offer a means to generate precise predictions that are carefully carried out in different ways for different circumstances (Fildes & Stekler, 2002). Forecasts serve a variety of purposes that are important to both the private and public sectors. These forecasts are used by the Federal Reserve to describe the health of the economy, identify issues, and justify monetary policy decisions (Federal Reserve System, 2016). Federal, state, and local governments also use these to justify their fiscal policy decisions (Hopper, 2018). In addition, industry sectors also make use of forecasts; the lumber industry has made use of many models to forecast lumber demand and to analyze shocks for a variety of situations.

Over time, the methods to develop forecasts have improved. Initially, basic least squared models were used such as two stage least squares (2SLS) and three stage least squares (3SLS) models, autoregressive (AR) and moving average (MA) models were also used, and then more precise vector autoregressive (VAR) models were used to generate macroeconomic forecasts; all these models assume that the data are linear and stationary (Tabor, 2017). Later, autoregressive integrated moving average (ARIMA), error correction (EC) and vector error correction (VEC) models were developed to account for cointegrated relationships since stationary models could not do so (Engle & Granger, 1987). Non-stationary (or cointegrated data) means the variables move with each other due to some relationship, and stationary data means variables don't move with each other due to a relationship (Lambert, 2013). It is important to test your data for stationarity, since using a model with the wrong data relationship would likely provide spurious results due to improper estimators (Hill et al., 2011). In addition, other hybrid or unique models such as models that include behavioral adaptive expectations have been developed (Haji-Othman, 1991). Stationarity can be tested by the Augmented Dickey-Fuller (ADF) test, and cointegration can be tested with the Johansen procedure (Al-Ballaa, 2005; Johansen, 1988). These tests should always be utilized to confirm whether the data is stationary or cointegrated (depending on the type of model that is being utilized).

Once the models are solved, the estimates can be applied to forecasting techniques that vary depending on the type of model being utilized. Forecasting is a critically important tool that can help project future forest uses and impacts for businesses, households in the area, and policy makers in the public and private sectors (Pattanayak et al., 2002). The "law of one price" has proven time series data for the U.S. and Canadian lumber market are non-stationary, which accordingly requires models that utilize cointegrated data (Yin & Baek, 2005; Song et al., 2011).

**Review of Stationary Models**

To address the simultaneous equation bias problem of supply and demand, 2SLS and 3SLS models were applied (Pattanayak et al., 2002):

$$2SLS: y_{it} = X_{it}\beta_i + \epsilon_t \ \& \ X_{it}\beta_i = \Gamma_{it}\alpha_i + Z_{it}\gamma_i$$

where $\Gamma$ are endogenous regressors, $Z$ are exogenous regressors, and $\epsilon_t$ is a white noise error term. 3SLS utilizes this framework but adjusts this to exploit correlation of disturbances across equations (McFadden, 1999).

But many studies with these approaches have not properly accounted for the complications of time series properties (Parajuli & Chang, 2015). To better account for time series properties, AR and MA models were introduced:

$$AR: y_t = \alpha_0 + \sum_{i=1}^{p} \alpha_i y_{t-i} + \epsilon_t$$

where $p$ is the order, $\alpha_1, \ldots, \alpha_p$ are the parameters of the model, $\epsilon_t$ is a white noise error term, and $\alpha_0$ is a constant.

$$MA: y_t = \sum_{i=0}^{q} \beta_i \epsilon_{t-i}$$

where $q$ is the order, $\beta_0, \ldots, \beta_p$ are the parameters of the model, and the $\epsilon_t, \ldots, \epsilon_{t-q}$ are white noise error terms. Autoregressive models show that output depends on its' previous values and an imperfectly predictable term (Jong, 2013). Moving average models show that output depends on current and various past values of an imperfectly predictable term (Jong, 2013). The stationary combination of AR and MA models are called ARMA models. VAR models are simply a vector of auto regressive models that can account for multiple variables (Hauser, 2018):

$$X_t = C + AX_{t-1} + \cdots + AX_{t-k} + \epsilon_t$$

where $X_t$ is a vector of variables, $k$ is the number of lags, $C$ is a vector of constants, $A$ is a vector of parameters, and $\epsilon_t$ is a vector of error terms. All these models are linear and require the data to be stationary (Tabor, 2017). If the data are nonstationary, the results would likely be spurious due to improper estimators (Hill et al., 2011).

Although most stationary forecasting models work with supply and demand for forest products, there are differences in the applications of different models to different markets. The forest products supply and demand market in Australia was analyzed via the 2SLS model by Ferguson (1973), and in the US by Song et al. (2011). Ferguson calculated long run price and income elasticities but faced major collinearity issues due to the limited number of observations. Song et al. also used the 2SLS model to determine long run elasticities but improved upon the basic 2SLS model by including dummies, trends, and up to 12 lags as instrumental variables.

Pattanayak et al. (2002) used a 3SLS model to analyze the southern region of the US, with a particular focus on ownership of forestry. Daigneault et al. (2016) used a 3SLS model to analyze multiple regions of the US for welfare analysis. This welfare analysis utilized short run and long run elasticities, demand, and supply. The southern region of the US was studied by Mei et al. (2010) with a VAR model. Zhou et Buongiorno (2006) used a specialized ARMA called STARMA so they could run shock analyses that utilize IRFs (impulse response functions). VAR models were also used for Finland's sawlog markets (Malaty et al., 2006; Hetemaki et al., 2004). Malaty et al. focused on short run stumpage price forecasts in different regions of Finland. Hetemaki et al. looked at German import demand as well as Finnish exports and demand. They specifically looked at how short-term changes in German lumber import demand affect forecasts for Finnish lumber exports, which then affect sawlog demand in Finland. Some of these specialized models found unique ways to deal with non-stationarity, but studies in the next subsection utilized models that were built to directly handle non-stationary data.

**Review of Cointegrated Models**

If data are not stationary, then cointegrated models need to be utilized such as ARIMA, EC or VEC models (Engle & Granger, 1987). Cointegration is when nonstationary series drift together instead of moving apart (Lambert, 2013). This means that even though two variables seem unrelated in the short run, they achieve equilibrium in the long run (Lambert, 2013). If $y_t$ and $x_t$ are nonstationary, but their linear combination $(y_t - \beta x_t)$ is stationary, then $y_t$ and $x_t$ are a cointegrated series (Lambert, 2013). Since the "law of one price" proved that the US and Canadian lumber markets are not stationary, many research (including this paper) focus on working with non-stationary time series models (Yin & Baek, 2005; Song et al., 2011). ARIMA models are the combination of AR and MA models; this is a linear model that accounts for cointegration due to its differencing term $d$ (Penn State, 2018):

$$(1-L)^d y_t = \alpha_0 + \sum_{i=1}^{p} \alpha_i \Delta y_{t-i} + \sum_{i=0}^{q} \beta_i \epsilon_{t-i}$$

where $L$ is the lag operator. EC models are as follows (Lambert, 2013):

$$EC: \Delta y_t = \delta_0 + \sum_{i=1}^{p} \delta_i \Delta x_{t-i} + \sum_{j=1}^{k} \mu_i \Delta y_{t-j} - \lambda(y_{t-1} - \alpha - \beta x_{t-1}) + \epsilon_t$$

where the one cointegrated relationship between $x$ and $y$ is $(y_{t-1} - \alpha - \beta x_{t-1})$. But if there are multiple cointegrating equations, a vector of EC models known as a VEC model should be used (Parajuli & Chang, 2015):

$$VEC: \Delta X_t = \sum_{i=1}^{k-1} \Gamma_i \Delta X_{t-i} + \Pi X_{t-1} + \mu + \phi D_t + \epsilon_t$$

where $X_t$ is a vector of variables, $k$ is the lag length, $\mu$ are constant terms, $\phi D_t$ measures the number of cointegrating equations, $\Pi \Delta X_{t-i}$ and $\Gamma X_{t-1}$ represent the lag structure, $\Pi$ are long run coefficients, and $\Gamma$ are short run coefficients.

In addition to utilizing a 2SLS model as we mentioned earlier, Song et al. (2011) also used an EC model to find short run elasticities. Song et al. (2012) argued that EC models should be utilized due to their relatively more robust results compared to alternative methods. This argument is also supported by Polemis (2007) in their analysis of energy demand in Greece, which argues for utilizing models built to handle cointegrated data. Mei et al. (2010) also used an EC model to analyze sawlog stumpage prices in the southern region of the US. Their purpose of utilizing multiple models was to compare forecasting accuracy, but they do not account for cointegration in this model (like they do for their EC model). Due to the limitation of only being able to utilize one cointegrated equation, there is a lot that has been done with VEC models instead to handle multiple cointegrated equations. Both Mei et al. (2010) and Malaty et al. (2006) also used an ARIMA model which was compared to the VAR model for forecasting accuracy. As well as utilizing a VAR model, Hetemaki et al. (2004) also utilized an ARIMA and VEC model since they found cointegration present. All of these models were estimated to compare forecasting accuracy, and it was found that VAR and VEC models were both similar and better than an ARIMA model.

Parajuli and Chang (2015) used a VEC model to estimate simultaneous supply and demand equations for the market in Louisiana. Parajuli et al. (2016) used Parajuli and Chang's 2015 data analysis to argue that using a VEC model compared with simultaneous equations utilizing a 2SLS model produced similar supply and demand coefficients. It is important to note that Parajuli et al. and Hetemaki et al. do not account for the "law of one price" when comparing their models, which is a big argument for using models that account for cointegration (Yin & Baek, 2005; Song et al., 2011). This can be seen by their use of stationary modeling for non-stationary data series. A VEC model was also used by Toppinen (1998) to model the Finnish market, specifically estimating short and long run coefficients of demand and supply models. Hietala et al. (2013) extended the analysis of the Finnish market by analyzing Swedish and Finnish export markets of sawlogs to the UK with a VEC model. This was done to analyze the effect of the founding of the European Monetary Union on trade flows. Baek (2012) estimated US lumber imports from Canada with a VEC model. They found that for softwood lumber trade US lumber prices and housing starts are more important compared to the bilateral exchange rate. Nagubadi and Zhang (2013) also used a VEC model to estimate US lumber imports from Canada and found that in the long run, the quantity of US lumber imports from Canada was positively affected by US domestic lumber prices and negatively affected by Canadian lumber prices. They used these findings to motivate that trade agreements between the US and Canada can have very important economic effects for both countries.

This Literature Review section has looked at the basics and applications of forecasting macroeconomic variables, the details of various stationary models and cointegrated models, and how prior studies have utilized these models to analyze the lumber market. It is important to note that we are looking at different but closely related industries, but the literature regularly uses "lumber" terminology which corresponds to what we reference as the forest products industry. Forecasts are most commonly performed for macroeconomic variables such as employment, output, prices from inflation, and the federal funds rate. Even though there are measurement errors when calculating forecasts, they are still used for a variety of important private and public sector analyses. Forecasts are primarily used for fiscal policy for governments, and monetary policy for the Fed. This paper utilizes a VEC model to measure bi-directional and cross directional relationships (compared to the single direction of a typical regression). Another reason we use a VEC is that it allows us to handle multiple cointegrated equations for the data within this analysis. Lastly, our VEC model is used so we can simultaneously measure supply and demand by estimating long run and short run parameters for our variables (Parajuli & Chang, 2015; Toppinen, 1998). Our model measures supply and demand, but primarily focuses on output, employment and price.

Other stationary models were considered, but due to the "law of one price" cointegrated models are better suited to handle data on the lumber industry in Canada and the US (Yin & Baek, 2005; Song et al., 2011). It is important to recall that nonstationary data utilizing stationary models would provide improper estimators which would likely lead to spurious results (Hill et al., 2011). Stationarity and cointegration can be tested by the ADF test and Johansen procedure (Al-Ballaa, 2005; Johansen, 1988). Other cointegrated models were considered, but since we expected more than one cointegrating relationship, the VEC model was deemed the most appropriate. In fact, even though prior literature on the lumber market utilized stationary models, these studies either analyzed outside of the US or Canada, compared these models to cointegrated models for comparison, accounted for cointegration by augmenting the model, or did not properly account for generally cointegrated lumber market data. Estimates from these models can be applied to forecasting techniques which is important for projecting future forest uses and impacts (Pattanayak et al., 2002). This paper's VEC model will then be used to provide forecasts that will be used to analyze key macroeconomic variables for Maine's forestry and logging industry. Our VEC model is used instead of a VAR to account for cointegration, and the VECs vector modeling allows us to measure bi-directional and cross directional relationships. This means we can see how everything impacts each other while accounting for everything.. Our VEC model is also used so we can simultaneously measure supply and demand by estimating long run and short run parameters for our variables (which has been done in literature) (Parajuli & Chang, 2015; Toppinen, 1998).

## Methodology

In this section we first discuss the data that will be analyzed in our model, the method that will be used to construct and estimate our model, what these estimates mean, as well as how these estimates will be utilized to perform forecasts of our data.

**Data**

Our data are made up of macroeconomic and related quarterly data from 2001Q1-2018Q1, Table 1 provides descriptions and sources for each variable.

Table 1. Variable List. Includes variable names, descriptions, and sources used to model Maine's forestry and logging industry.

| Variable | Description | Source |
|---|---|---|
| Employment | Employment in Maine for all NAICS 113 (in thousands) | BLS |
| Number of Firms | Employment in Maine for private NAICS 113 | BLS |
| Exchange Rate | Canadian to US Exchange Rate (not seasonally adjusted) | Board of Governors of the Federal Reserve System (US) |
| Price | Producer Price Index for Lumber and Wood Products in 2017 dollars (NAICS 321, not seasonally adjusted) | BLS |
| Wages | Maine's total wage for private NAICS 113 (in thousands) | BLS |
| Output | Proxy of Maine's RGDP (not seasonally adjusted, in millions) | BEA |

Employment, the number of firms, and wages are for the forestry and logging industry which is measured as NAICS code 113. All these variables account for supply and demand of this industry in Maine. Output, wages, number of firms, and price impact supply while price, employment and exchange rate impact demand. Total wages were used instead of wages per capita since wages per capita would assume that everyone in the forestry and logging industry makes the same wage, which would be a strong assumption. Price instead measures wood products (which is measured as NAICS 321). Quarterly output was proxied from quarterly US real gross domestic product (RGDP) and annual Maine RGDP data. We then used the percentage of US RGDP that consisted of Maine's RGDP to adjust Maine's annual RGDP to quarterly RGDP. This assumes other industries that are included within GDP haven't significantly changed, so we are only seeing changes in the forestry and logging industry. Table 2 provides summary statistics of the variables used in this paper; these summary statistics show that there are no missing observations and each variable has reasonable values, range, and standard deviation. This can also be seen in Figure 4 which shows graphs of these variables over time.

Table 2. Summary Statistics. Includes summary statistics of key variables used to model Maine's forestry and logging industry.

| Variable | Mean | Standard Deviation | Minimum | Maximum | N |
|---|---|---|---|---|---|
| Quarter | n/a | n/a | 2001Q1 | 2018Q1 | 69 |
| Employment (in thousands) | 2.513 | 0.164 | 2.067 | 2.800 | 69 |
| Wages | 22817.350 | 4246.641 | 13490.000 | 28600.000 | 69 |
| Exchange Rate | 1.207 | 0.183 | 0.968 | 1.595 | 69 |
| Number of Firms | 467.652 | 31.011 | 420.000 | 539.000 | 69 |
| Price* | 0.852 | 0.096 | 0.665 | 1.078 | 69 |
| Output* | 13408.380 | 397.173 | 12354.850 | 14258.370 | 69 |
| *The following data is cointegrated | | | | | |

Figure 4. Variable Graphs. Graphs of all variables we used from their time span of 2001Q1-2018Q1.

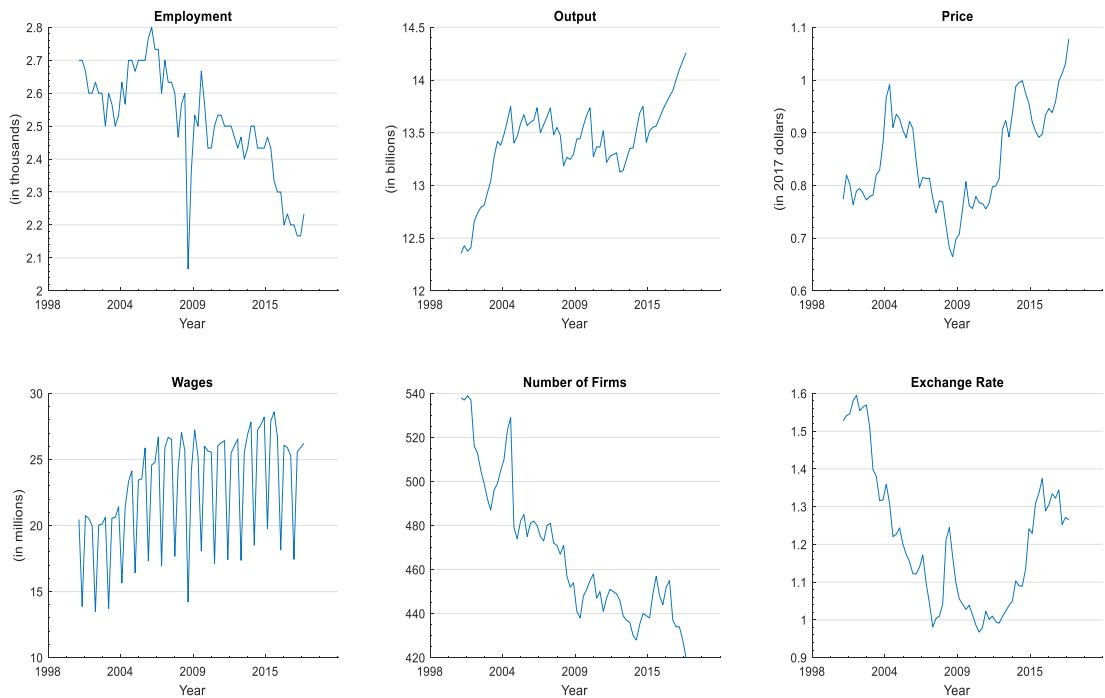

**Model Construction & Justification**

We analyze our six variables (employment, output, price, wages, firms, and exchange rate) as a system of equations in a 6<sup>th</sup> dimensional VEC model as:

$$\Delta X_t = \sum_{i=1}^{k-1} \Gamma_i \Delta X_{t-i} + \Pi X_{t-1} + \mu + \phi D_t + \epsilon_t$$

$$X_t = [output_t, price_t, employment_t, wage_t, exchangeRate_t, numberFirms_t]'$$

$$t = 2001Q1 - 2018Q1$$

where $X_t$ is a vector of our variables, $k$ is the lag length, $\mu$ is a vector of constant terms, $\phi D_t$ is a vector that measures the number of cointegrating equations (aka the rank), $\Pi \Delta X_{t-i}$ and $\Gamma X_{t-1}$ are vectors that represent the lag structure, $\Pi$ are long run coefficients, and $\Gamma$ are short run coefficients. The $\Delta X_t$ estimates measure the impact of everything used to analyze the impact of two variables in an IRF, and $\mu$ is used for the Johansen restriction. The IRFs are then utilized to perform forecasting. While output, price, and employment forecasts are the focus of this research, we also forecast other macroeconomic and related variables.

To utilize this model, we need to test for cointegration and lag structure. By using Johansen's test for cointegration and lag tests, our data are found to be cointegrated at rank 2 and lagged at rank 4 (see Table 3 and Table 4). The lag tests we used were Akaike Information Criterion (AIC), Final Prediction Error (FPE), and sequential likelihood-ratio (LR); these were most appropriate due to our smaller sample size (Liew, 2004).

Table 3. Lag Tests. Performed for our VEC model, we focused on AIC, FPE, and LR.

| Lag Tests ||||||||
| --- | --- | --- | --- | --- | --- | --- | --- |
| Sample: 2002Q1-2018Q1 ||||||| Number of Observations: 65 |
| Lag | Tests ||||| D.f. | P-Value |
| | LL | LR | FPE | AIC | HQIC | SBIC | | |
| 0 | -1223.9 | | 1.10E+09 | 37.84 | 37.92 | 38.04 | | |
| 1 | -919.13 | 609.49 | 282,415.00 | 29.57 | 30.13 | 30.98* | 36 | 0 |
| 2 | -850.65 | 136.97 | 106,872.00 | 28.57 | 29.61 | 31.18 | 36 | 0 |
| 3 | -805.93 | 89.44 | 87,868.20 | 28.31 | 29.81 | 32.12 | 36 | 0 |
| 4 | -699.43 | 213.00* | 11,604.20* | 26.14* | 28.12* | 31.15 | 36 | 0 |
| * = Number of Suggested Lags ||||||||
| *Note: Everything is held endogenous except for a constant term* ||||||||

Table 4. Rank Test. Johansen test performed for our VEC model to measure rank.

| Johansen Tests for Cointegration ||||||
| --- | --- | --- | --- | --- | --- |
| Maximum Rank | Parameters | Log Likelihood | Eigenvalue | Trace Statistic | 5% Critical Value |
| 0 | 42 | -964.248 | | 171.843 | 94.150 |
| 1 | 53 | -918.684 | 0.743 | 80.713 | 68.520 |
| 2 | 62 | -898.412 | 0.454 | 40.169* | 47.210 |
| 3 | 69 | -887.281 | 0.283 | 17.907 | 29.680 |
| 4 | 74 | -882.707 | 0.128 | 8.761 | 15.410 |
| 5 | 77 | -878.396 | 0.121 | 0.137 | 3.760 |
| 6 | 78 | -878.327 | 0.002 | | |
| * = Number of Cointegrating Equations ||||||
| Trend: Constant ||| Number of Observations: 67 |||
| Sample: 2001Q3-2018Q1 ||| Number of Lags = 2 |||

It is also important to test for autocorrelation, normality, and cointegration specification which are shown in Tables 5, Table 6 and Figure 5.

Table 5. Residual Autocorrelation Test. Lagrange-multiplier test performed for our VEC model to measure autocorrelation.

| Lagrange-Multiplier Test ||||
| --- | --- | --- | --- |
| Lag | $\chi^2$ | D.f | P-Value |
| 1 | 33.8489 | 36 | 0.5713 |
| 2 | 46.3709 | 36 | 0.1154 |

Table 6. Normality Tests. Performed for our VEC model, these include Jarque-Bera, Skewness, and Kurtosis.

| Normality Tests | | | | | | | | | | |
|---|---|---|---|---|---|---|---|---|---|---|
| Test Type | Jarque-Bera Test | | | Skewness | | | | Kurtosis | | |
| Equation | $\chi^2$ | D.f. | P-value | Skewness | $\chi^2$ | D.f. | P-value | Kurtosis | $\chi^2$ | D.f. | P-value |
| D_output | 9.075 | 2 | 0.011 | -0.508 | 2.796 | 1 | 0.095 | 4.523 | 6.280 | 1 | 0.012 |
| D_price | 5.445 | 2 | 0.066 | -0.374 | 1.514 | 1 | 0.218 | 4.205 | 3.931 | 1 | 0.047 |
| D_employment | 13.654 | 2 | 0.001 | -0.553 | 3.308 | 1 | 0.069 | 4.955 | 10.346 | 1 | 0.001 |
| D_wages | 1.226 | 2 | 0.542 | 0.133 | 0.191 | 1 | 0.662 | 3.618 | 1.035 | 1 | 0.309 |
| D_exchangeRate | 0.007 | 2 | 0.997 | -0.007 | 0.001 | 1 | 0.982 | 2.952 | 0.006 | 1 | 0.937 |
| D_numFirms | 2.308 | 2 | 0.315 | -0.308 | 1.030 | 1 | 0.310 | 3.687 | 1.278 | 1 | 0.258 |
| ALL | 31.715 | 12 | 0.002 | | 8.839 | 6 | 0.183 | | 22.875 | 6 | 0.001 |

Figure 5. Cointegrating Equation Specification Test. Performed for our VEC model, this ensures we properly accounted for cointegration.

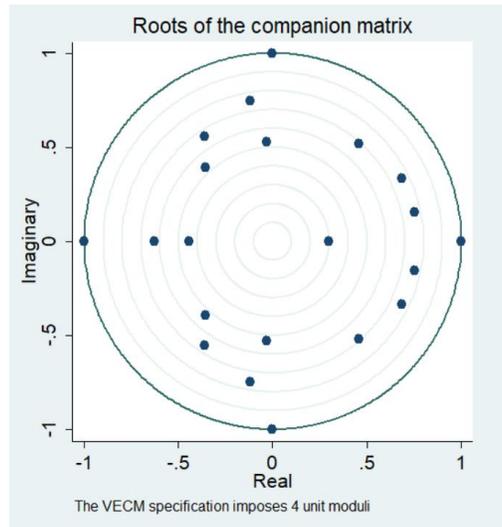

Normality tests (including Jarque-Bera, Kurtosis, and Skewness), the eigenvalue stability test, and the Lagrange-Multiplier test for autocorrelation show that the cointegrating equations are properly specified, residuals are not autocorrelated (at the lag of 2), and errors are not normally distributed for wages, exchange rate, and the number of firms. Since this is cointegrated data, if errors are not normally distributed, there is no need for concern. Now that the model is properly constructed and tested, we will further discuss how the estimates will be utilized for forecasting.

**Forecasting**

Impulse response functions measure the impact that one variable has on another variable, more precisely the reactions of variables to shocks hitting the system. So if there was a change in price for instance, we would then be able to see how each of our variables would respond. These IRFs are then utilized to forecast variables into the future. As mentioned earlier, while this research is primarily interested in output, price and employment forecasts, other variables are also important to forecast. As we saw in the Literature Review section, a variety of other literature for lumber markets have forecasted supply and demand for lumber in a variety of regions and countries.

## Results

This section reports the forecasts for each of our variables, explains what they mean and assesses whether the forecasts are appropriate (based on actual trends). Our VEC model provides us with IRF estimates which can be seen graphically in Figures 7-10. These IRFs are used to perform forecasts for each of our variables. We forecasted each of our variables from 2018Q1 to 2023Q1, these results can be seen in Figure 6.

Figure 6. Variable Forecasts. Out of sample forecasts projected for our variables.

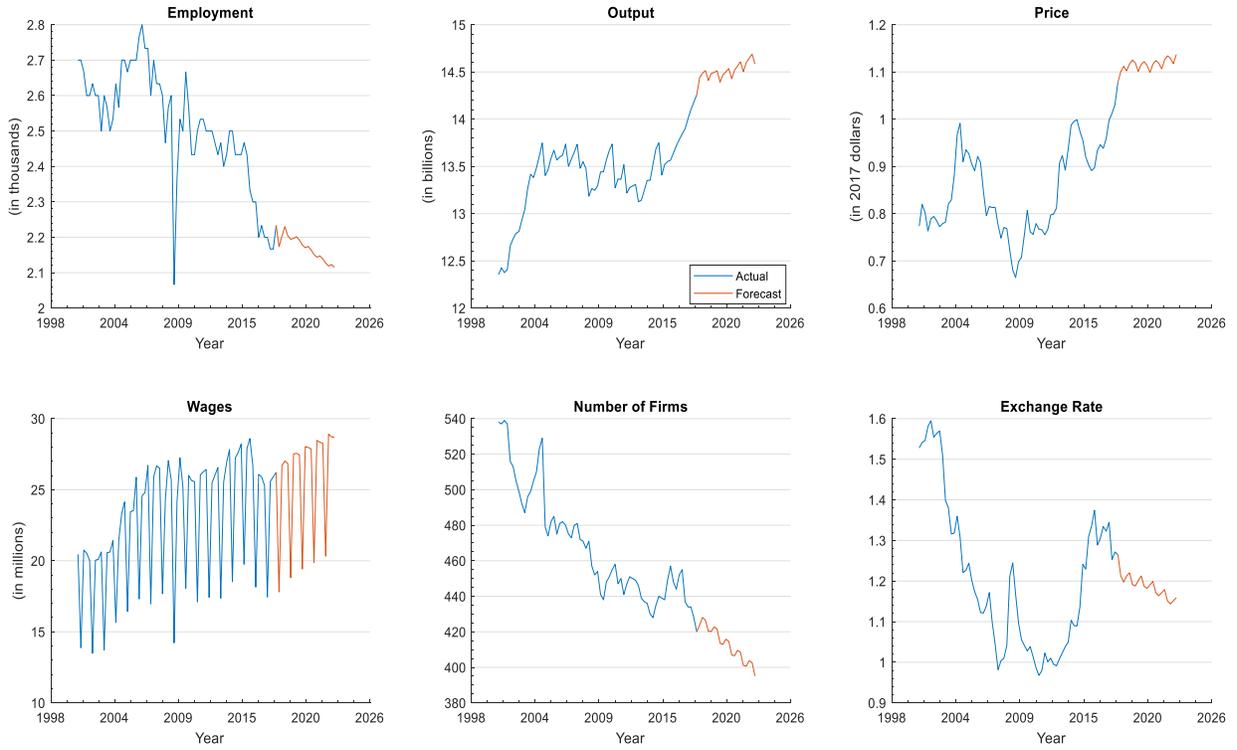

Figure 7. IRF for Employment. Orthogonalized IRF generated to support the response of wages to a shock in employment.

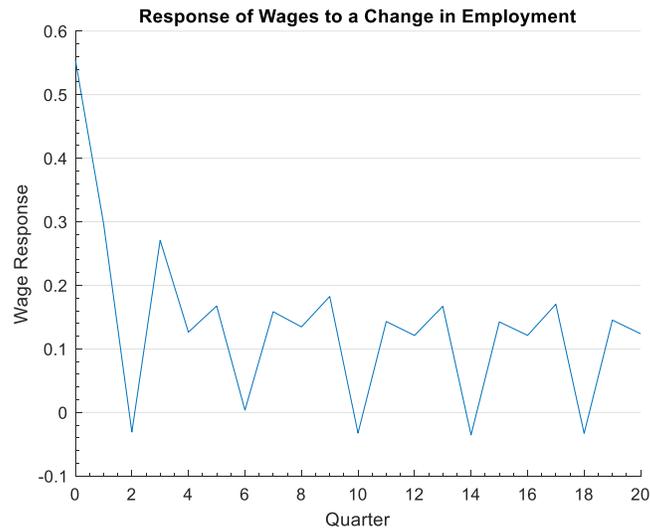

Figure 8. IRFs for Firms. Orthogonalized IRFs generated to support the responses of employment to a shock in the number of firms.

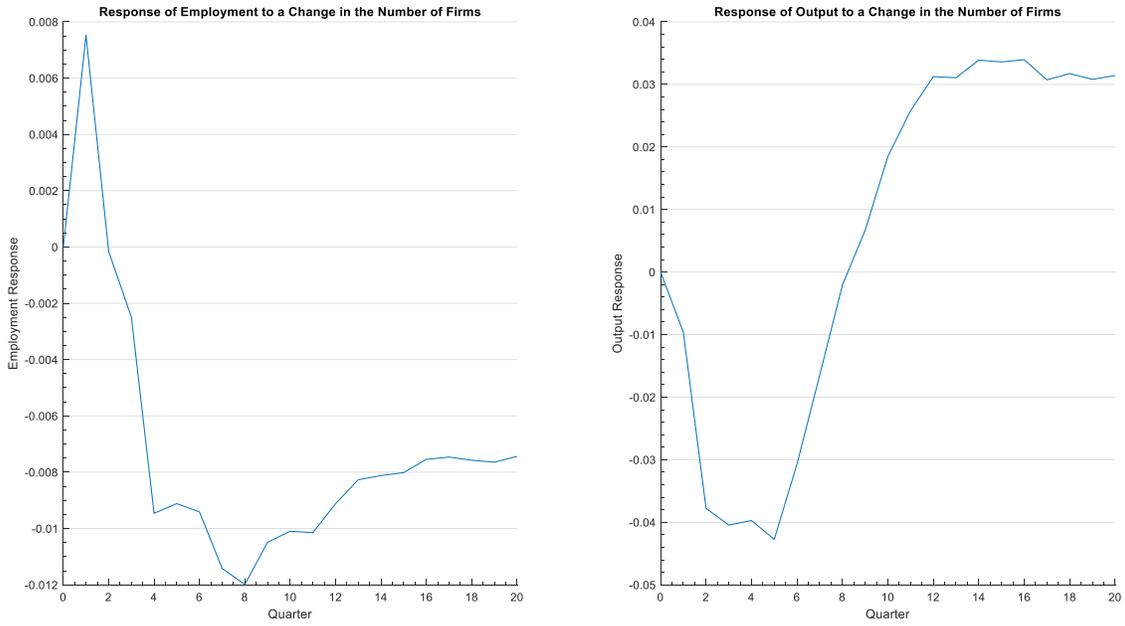

Figure 9. IRFs for Output. Orthogonalized IRF generated to support the responses of the number of firms to a shock in output.

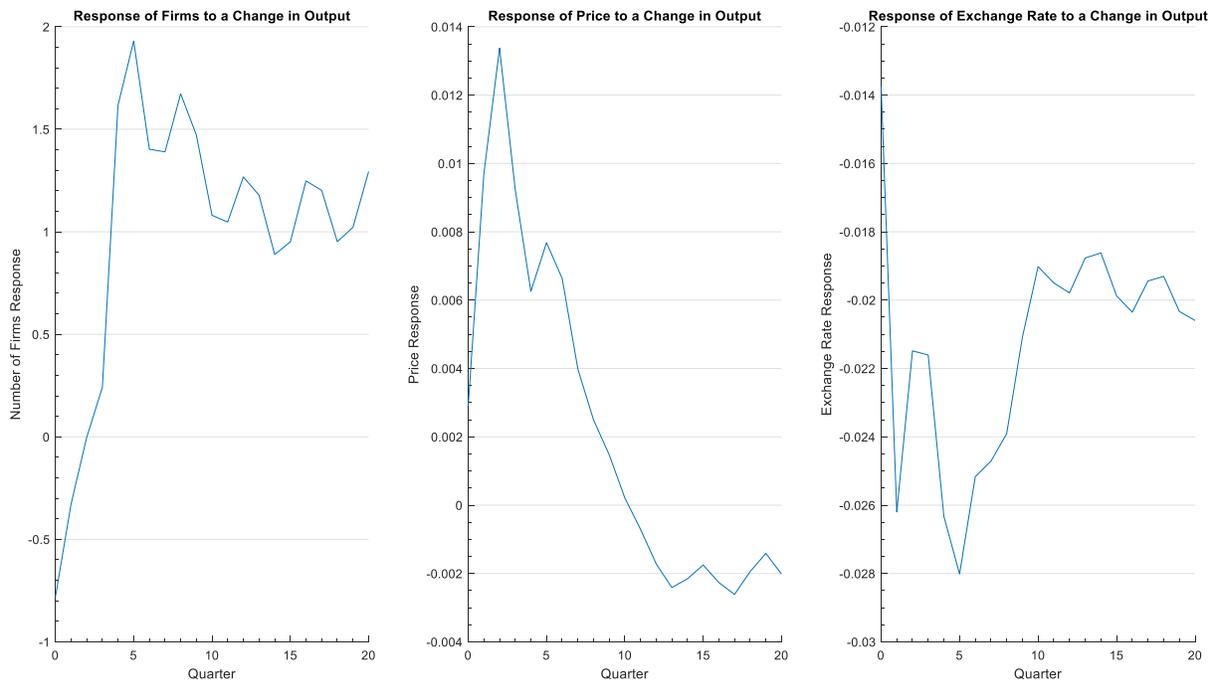

Figure 10. IRFs for Price. Orthogonalized IRF generated to support the responses of employment to a shock in price.

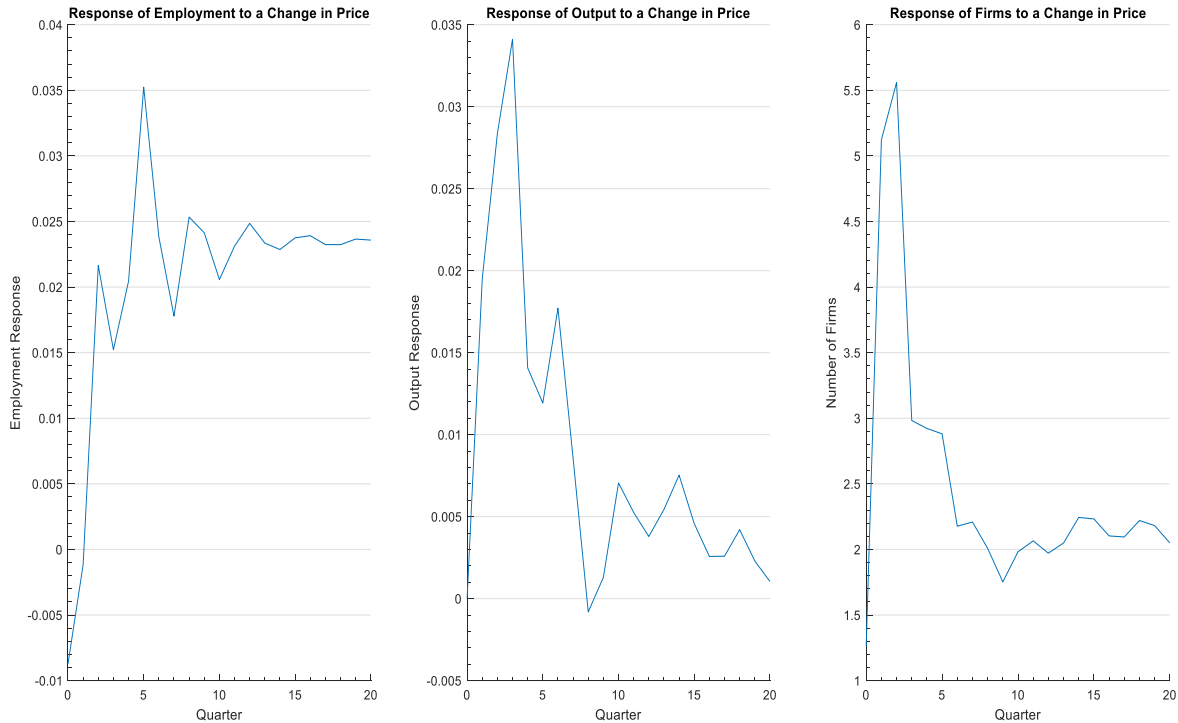
,

These resulting forecasts for Maine suggest that in 5 years output will slightly increase, employment will fall, prices will stay level, the number of firms will decrease, wages will cyclically increase slightly, and exchange rate will cyclically decrease slightly. All of these findings have important implications for the forestry and logging industry in Maine and impacts to the economy.

**Discussion**

These forecasts have important implications for Maine's forestry and logging industry as well as local economies. In summary, the forecasts imply that although the contribution of the industry will remain stable due to level prices and a slight increase in output for Maine, the local communities could be worse off due to lowered employment and reductions in the number of businesses in Maine. Based on these findings, rural communities in Maine (especially ones with tax bases dependent on mills) should focus on increasing their employment and business activity to offset this negative forecast. It is important to note that output is a proxy of Maine's RGDP, forest products in 2016 only made up 5% of Maine's GDP and forestry and logging is 1/10$^{th}$ of that 5%. Another option for the purpose of the general economy and well-being of these rural communities, is to find employment and business activity in other industries. Beyond this simple summary, the highlights of our forecasts that we will discuss in more detail are employment, output, and exchange rate. We focus on exchange rate instead of price since this adjusts in the short run (while price can only adjust in the long run according to macroeconomic theory).

According to our forecast, current employment in Maine will increase, which will be followed by a general decrease. This short-term increase makes sense since some shut down mills have been repurposed, and new types of lumber production have required more logs to operate (which need to be harvested and transported by the forestry and logging industry) (Crandall et al., 2017). For example, the mills in Old Town and Rumford Maine are reopening later this year, so employment will accordingly increase (Pendharkar, 2019). Our forecasts also show that recent output in Maine will increase, which will be followed by a lesser cyclical increase. This makes sense due to the same repurposing and new types of production mentioned earlier (which require more logs for input), as well as the steady increase of harvest volume due to forest landowners of the north properly managing their forests (USDA, 2011). According to our forecast, the recent exchange rate will decrease followed by a lesser cyclical decrease. This is different from the small cyclical increase in exchange rate that has been observed since 2018 by the Federal Reserve (FRED, 2019). It is important to note that our data ends at 2018Q1, and at this time exchange rate was slightly falling. Also, between 2015-2018 the exchange rate has been going up and down (even with seasonality accounted for), which could be due to changes in political trade negotiations. Therefore, it is not alarming that our exchange rate forecast does not match the actual recent trends of exchange rate.

Since these forecasts seem to appropriately model indicators, this paper has now offered a viable approach to modeling the forestry and logging industry in Maine. These forecasts and their implications are very important to consider for maintaining the health of the forestry and logging industry, as well as the connected communities (especially the rural communities).

**Limitations**

Although these forecasts provide highly beneficial insights, it is important to note their limitations. Most of Maine's data was only provided annually, but annual data did not provide enough observations to perform robust forecasts. Utilizing quarterly data helped us attain more observations, but we had to proxy output based on Maine's annual RGDP and quarterly US RGDP. We were also not able to provide specific quarterly stumpage prices, so we instead used the 2017 PPI of lumber and wood products. This unfortunately captured a different sector than forestry and logging, but this was the best available data for a closely related sector. Capital wasn't included either, so we cannot discuss how changes in capital affect mill operation (which could affect log demand from the forestry and logging industry). This misses the key argument of the balance between labor and capital, and how improvements in capital via automation and efficiency significantly changed demand. It is also important to mention that our sample only had 69 observations, which is not ideal in terms of robust econometric forecasting. Further research would be highly beneficial when more regional data regarding supply and demand become readily available, especially utilizing cointegrated modeling to satisfy the "law of one price" as discussed in the Literature Review section (Yin & Baek, 2005; Song et al., 2011).

**Advantages**

        We modeled the forestry and logging industry using the best data and methods currently available. This research utilizes a VEC model which is most appropriate for data with multiple cointegrated relationships. The "law of one price" shows cointegrated models should always be utilized for forestry and logging data in the US (Yin & Baek, 2005; Song et al., 2011). Estimates of the VEC model provide IRFs, which measure the impact that one variable has on another variable. We didn't use a basic EC model since we expected more than one cointegrating relationship, and a VEC model can easily analyze cross directional impacts (which EC models cannot). To measure accuracy of the forecasts, we compared actual to forecasted values for 2016 and 2017 for all of our variables (which can be seen in Figure 11). With the exception of price (which is an index for a different industry), all the other forecasts match very closely with the actual values. This shows that our forecasts very closely match the actual trends of our variables, which means these forecasts are very close to reality.

Figure 11. Robustness Checks for Variable Forecasts. Compares actual to forecasted values for 2016 and 2017.

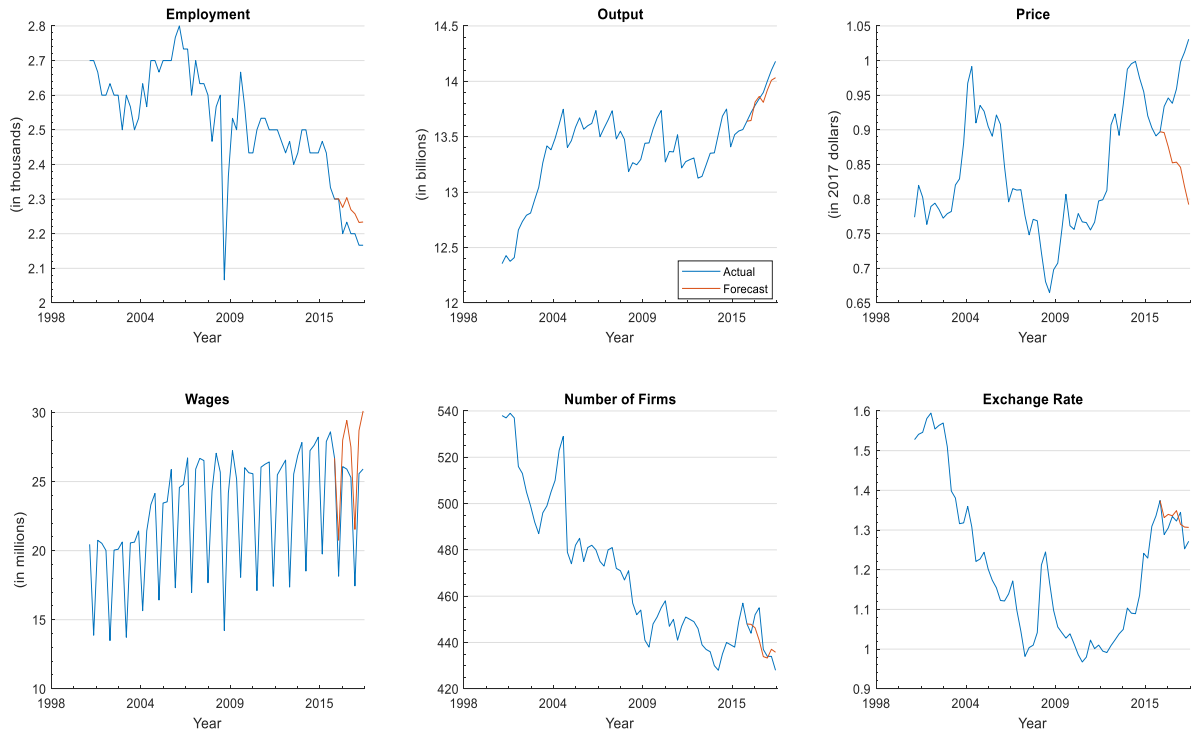

## Conclusion

One out of twenty-four jobs in Maine depend on the forest products industry, which has been devastated by the closing of 64% of the pulp and paper mills (Forest Opportunity Roadmap Maine, 2018; Crandall et al., 2017). The objective of this paper was to forecast key macroeconomic variables for this industry in Maine. Our key macroeconomic and related variables for this industry in Maine include industry employment in Maine, industry wages in Maine, number of Maine firms in the industry, output (measured as Maine's real gross domestic product), prices for US lumber and wood products, and the impact of the Canadian to US exchange rate on Maine's industry. These different variables account for supply and demand of the forestry and logging industry. Prior literature made use of various stationary and cointegrated modeling techniques for various lumber markets around the world, but since the "law of one price" has proven U.S. lumber market data is non-stationary, we utilized cointegrated modeling (Yin & Baek, 2005; Song et al., 2011).

Our analysis utilizes a VEC model to model multiple cointegrated relationships for quarterly data from 2001Q1 to 2018Q1 and forecasts up until 2023Q1 by making use of IRFs. Most available data for Maine was only available in annual terms, but we used quarterly data since there were not enough annual data observations to perform forecasts. Although we were not able to include specific stumpage prices or capital, this research still provides a significant contribution to research for the forestry and logging industry since it uses the best data and methods currently available, especially compared to other literature that does not properly account for cointegration in their modeling techniques. A VEC model does not just do well handling multiple cointegrating equations, but this type of model benefits from the ability to analyze cross directional impacts (compared to the other cointegrated modeling techniques). The forecasts show for the next five years in Maine output will slightly increase, employment will fall, prices will stay level, and the number of firms will decrease. Lowered employment and business in Maine imply the local community will be worse off, and level prices and a slight increase in output imply the industry itself will remain stable. These forecasts answer our 1st research question: How will key macroeconomic and related variables for Maine's forestry & logging industry change in the future?

For Maine's forestry and logging industry, these forecasts and analyses are especially important. Both the forestry and logging industry and forest products industry have struggled during the past 20 years with mill shutdowns and the remaining open mills have faced changes in demand. This is very important for towns in Maine that depend on employment in growing, harvesting, transport, support staff, and local mills. Policymakers and mill owners (especially in Maine) should take note of these projections to inform decisions they make and understand the implications they will have on the general economy, and more specific local economies. Our paper analyzed the forestry and logging industry in Maine, but this type of analysis could also be used for different industries, states or regions across the country. We encourage future studies in Maine when more data is available, and in other places where both the forestry and logging industry as well as the forest products industry are found to be important in the literature (such as Canada, Germany, Sweden, and Finland) (Hetemaki et al, 2004; Malaty et al, 2006; Zhang & Parajuli, 2016; National Association of Home Builders, 2017).

# CHAPTER 3: MAINE'S FORESTRY AND LOGGING INDUSTRY: ANALYZING THE IMPACT OF TARIFFS

## Introduction

Like any regional economy, Maine's economy has changed as industries internationally, nationally, and locally have changed over time. The forestry and logging industry over time has still been very important to Maine's economy. Some towns have been devastated by reduced demand for forest products as well as mill shutdowns. There are even still towns with operating mills that depend on them for employment of mill workers, harvesters, loggers, and support staff (Weeks, 1990). The forestry and logging industry includes the processes of growing trees, harvesting logs, transporting logs directly to buyers or processing mills (that convert logs to an end product), and the support staff throughout these stages of harvest and transport (Forest Opportunity Roadmap Maine, 2018). We will also use the term forest products industry to describe the sector that processes logs into lumber or other end products in a mill. For these towns, employment and related employment is at stake in addition to general business activity and even the tax base. During 2001, five towns had mills that made up more than 60% of the local tax base (Irland, 2001). This means that if the mill closed, the town would lose not just a lot of employment, but also most of its' tax base. In this paper, we will examine the history of this industry in Maine to see how these and other changes in the world have affected the industry and connected local communities. The objective of this paper is to analyze how a shock to the Canadian to US exchange rate (from an increase in tariffs) could affect employment and output.

## Literature Review

**Background of Tariff Changes**

There is a long-standing trade history between the US and Canada in lumber markets. This trade annually contributes $4 to $7 billion of goods, so tariffs and other trade restrictions could have large impacts to either or both countries (Zhang & Parajuli, 2016). Tariffs are a common trade restriction that changes the real cost of buying or selling goods, and retaliatory tariffs (new tariffs in response to the original tariffs in retaliation) are very plausible (Francois et al., 2018). About 48% of US softwood lumber imports are provided by Canada and 94% of Canadian softwood lumber imports are for the US, so there is a critical dependence between the two countries (ForestEdge LLC & Wood Resources Int LLC, 2018). Wear and Murray (2004) found that supply shocks between Canada and the US shrunk the timber supply by about 15%. They also argued that there is a constantly changing pattern of bilateral lumber trade between the two countries, so it is critically important to properly understand the market and macroeconomic factors (Baek, 2012).

While our research focuses on effects of shocks from the Canadian to US exchange rate on employment and output, we can learn from the other analyses and roles of exchange rates. Examples include utilizing the exchange rate with regards to international trade flows for the forestry and logging industry (Hietala et al., 2013). The exchange rate can be used as a parameter; when the exchange rate appreciates, US firms have a harder time exporting, and US prices of imports decline (Irland, 2017). Previous studies have estimated the elasticity of exports to changes in exchange rates (Adams et al., 1986; Bolkesjo & Buongiorno, 2006). Sun and Zhang (2003) analyzed how uncertainty of exchange rates can negatively affect trade volumes. This means that uncertainty in trade could negatively affect trade directly or indirectly through expectations. Hietala et al. (2013) argue that exchange rate changes have been used to gain price advantages against competitors.

Tariffs and other trade restrictions have been found to have a significant negative impact on US imports from Canada and a positive impact on imports from the rest of the world (Nagubadi & Zhang, 2013). Zhang and Parajuli (2016) show that historic Canadian lumber exports with the US prove that tariffs and other trade restrictions certainly have an effect on the number of exports. Figure 12 shows the historic pattern of Canadian lumber exports and shows the impact of tariffs and other trade restrictions.

Figure 12. Canadian Exports Graph. Historic graph of Canadian lumber exports with 4 periods of tariffs and trade restrictions highlighted (Zhang & Parajuli, 2016).

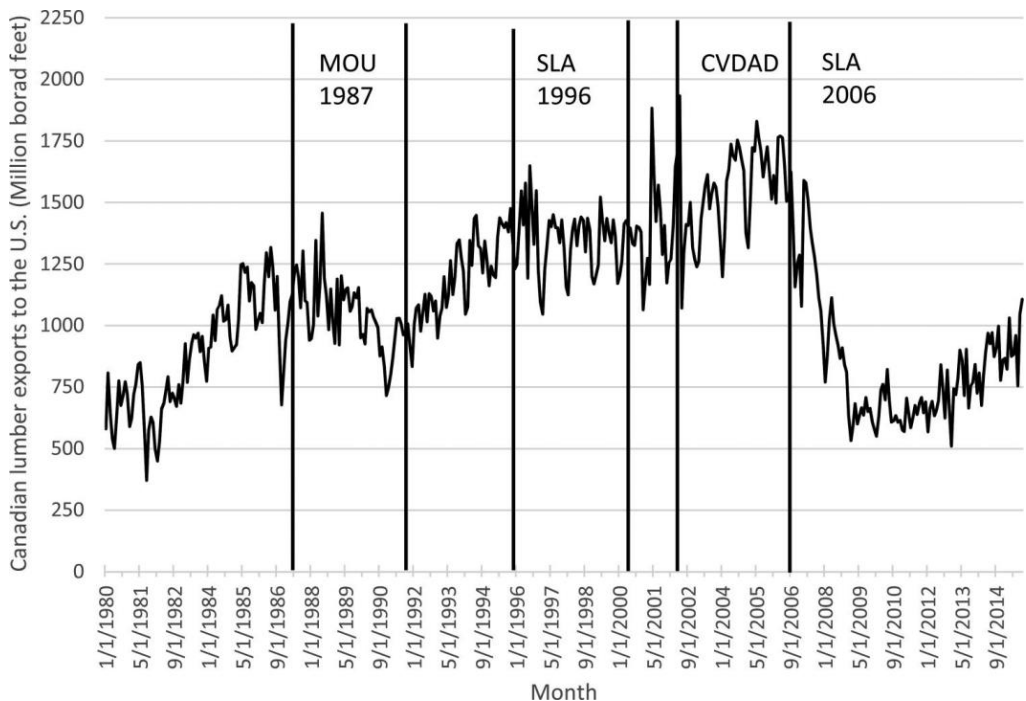

Tariffs between the US and Canada have averaged between 15%-20%, with 4 different periods of trade with the US. Trade restrictions had the largest range of 10%-28% when the US implemented CVDAD on Canada from August 2001 to September 2006 (Zhang & Parajuli, 2016). This was the largest impact and reduced US imports by 13%. US and Canadian softwood lumber are not perfect substitutes because the species composition is not the same, and consumers view imported lumber as a different good compared to domestic lumber (Buongiorno, 1979). If Canadian import prices were excluded, this could result in biased estimates from an omitted variable bias (Wooldridge, 2006). Our analysis is focused on direct domestic effects and indirect foreign effects, so we won't run into this problem.

Most recently the US government enacted various tariffs in various industries causing serious concerns. Look at the steel and aluminum tariffs implemented in 2017 for instance; even though some American jobs have been rekindled, US jobs overall have been lost. It has been projected that these tariffs caused 16 jobs to be lost per 1 job gained, which means 402,445 jobs were lost nationwide and 1,948 jobs were lost in Maine (Francois et al., 2018). President Clintons' economic advisor Dr. Laura Tyson explained this would occur due to potential retaliation tariffs, higher input prices, and global trade slowing down (Tyson, 2019). These tariffs impact any services or production that makes use of steel or aluminum which includes heavy machinery, cars, construction, motorbikes, food, and drinks (Costa, 2018). This example of an unrelated yet recent tariff enacted by the US government shows how there is a much deeper impact than simply the initial tariffs.

Literature has gone beyond basic statistics by using a computable general equilibrium model (Francois, 2018; Ciuriak & Xiao, 2018). This was done to model the entire US economy and assess impacts of policy shocks, but this was not computed at any regional levels. Ciuriak and Xiao (2018) found this would cause severe difficulties for individual firms and projects due to supplier-customer relationships. The International Monetary Fund (2018) argue that knock on effects from the negative responses of financial markets would occur due to rising trade tensions and conflicts, geopolitical concerns, and mounting political uncertainty. A global simulation model was used to also measure China's potential retaliatory tariffs on agricultural commodities including soybeans, cotton, sorghum, and pork (Zheng, Wood, Wang, & Jones, 2018). These are not unrelated from steel and aluminum tariffs, since these are China's potential retaliatory tariffs in response to the US recently implementing steel and aluminum tariffs. This model also doesn't analyze anything at the regional level, but simultaneously assesses trade policy changes at the industry level, national level, and global level. This study finds that US soybean producers would loose $1.8 billion between 2018 to 2019 in result of China's potential retaliatory tariffs. Since literature on the recent steel and aluminum tariffs and trade restrictions have only modeled the potential impacts nationally and internationally, regional analysis of the impacts of tariffs and trade restrictions for these goods and other would be a beneficial area for further study.

Overall, US businesses paid $375 million in tariffs during November of 2017; these businesses paid more than 7 times ($2.7 billion) in tariffs during November of 2018 (Tariffs Hurt the Heartland, 2019). This caused exports for the US to fall by 37%, which severely decreased output. Francois et al. (2018) show this $36.8 billion estimated decrease in US GDP will severely hurt the economy. Oxfords' lead economist Adam Slater found that the proposed tariffs will increase over 13 times to $800 billion (or 4% of world trade) (Edwards, 2018). Both the forestry and logging industry as well as the forest products sector have seen similar negative impacts. The National Association of Home Builders (2017) found US lumber prices jumped by 22% simply in anticipation of new tariffs and trade restrictions with Canada. This increase caused housing prices to rise, cut $350 million in taxes for local governments, cut 8,241 jobs, and cut $598 million in wages. Increases in tariffs for the forestry and logging industry and the related forest products industry, other industries, and the entire country could have negative effects for the specific industries. These increases could also set off a chain of events for interrelated businesses and communities.

**Background of Mill Changes**

Both the forestry and logging industry and forest products industry began in Maine during 1634, and Bangor Maine became the worlds' largest lumber shipping port by 1830 (Judd et al., 2010). In the past few decades logging moved away to the Northwest and South; at that time, Maine also started to face competition from other regions of the US as well as Canadian lumber suppliers. Offshoring has also caused US output overall to decrease, and the Great Recession further decreased US output. Both the forestry and logging industry and the forest products industry collapsed during the Great Recession due to the collapse of housing construction in 2006 (Woodall et al., 2011). In addition, automation in the 1990s led to a general decrease in demand for manufacturing employment and globalization decreased US manufacturing output (Irland, 2017). The rise of electronic media (such as e-mail and social media) since the mid-1990s has lowered demand for print media (such as letters and newspapers). The largest decrease was North American newsprint which declined 55% between 2000-2014 (Irland, 2017). In contrast to print media's demand decrease, packaging demand has increased due to shipping services (Berg & Lingqvist, 2017). Even though Maine is no longer a national leader for both the forestry and logging industry and the forest products industry, it still relies on them (Smith et al., 2009).

Demand for the forest products industry stabilized in the long run, but these shocking transitions still caused over half of the mills in Maine to shut down (Weeks, 1990). This has made businesses and people suffer, especially those in rural mill towns. Irland (2017) found the direct effect of a single mill shutdown could easily cost 100 jobs or more, but this doesn't account for local food service and lodging businesses that service these mill employees and their families. It was also found that property prices fell, rental vacancy rates declined, and tax valuations declined. To give an idea of what the tax contribution of mills looked like, Maine in 2001 had 5 towns where the mills made up 60% or more of the local tax base (Irland, 2001). One mill provided as much as 85% of the tax revenue of the town. Local hospitals also depended on the health insurance mill workers had in unionized mills, and local infrastructure investments were left stranded. Even just cleanup and redevelopment of mills is very costly, this doesn't include the tax debt most of these closed down mills hold (Irland, 2017).

In the past 20 years, 11 out of 17 mills closed which resulted in Maine losing 73% of its' paper jobs. This is similar to Wisconsin and Oregon that both lost 60%, but Georgia in the Southeast only lost 41% due to its' more modern mills with different products. After the recession, 7 mill closures occurred which shows the industry still struggled even when economic conditions improved (Irland, 2017). Irland also argues that during the past 20 years, the appreciation and depreciation of the exchange rate contributed to many of the mill shutdowns. Just like mill shutdowns have negative multiplier effects (via direct and indirect effects), mill openings have positive multiplier effects. For instance, the mills in Old Town and Rumford Maine are being repurposed by a Chinese firm. This firm plans to invest $100 million for construction and also to create over 100 jobs for the mill in Old Town (Pendharkar, 2019). The local construction workers and newly hired mill workers will now have new income that they will use to pay for goods and services in the area. This shows that the positive effect is larger than just the initial investment and job creation.

**Background of Demand Changes**

    Lumber is used in construction (primarily for homes), wood fuels, and furniture, so the industry is affected when demand for any of these change (U.S. Congress, 1983). Many changes in demand for forest products have caused many mills to shut down, and many of the surviving mills faced significantly less demand. To keep both the forestry and logging industry and forest products industry afloat, many mergers occurred which concentrated the industry (Irland, 2017). But even this improvement could not overcome the effects of shrinking demand for forest products. Mills in the south are more efficient than mills in Maine and are also closer to end users, so the south has a competitive advantage over Maine (Irland, 2017). This history for both the forestry and logging industry as well as the forest products industry showcase what has changed demand for forest products over time. Figures 13-15 help visualize how US output, Maine employment, and the US housing market have changed over time.

Figure 13. US Single-Family Housing Starts & Home Ownership Graph. This graph shows important markets for the US housing market.

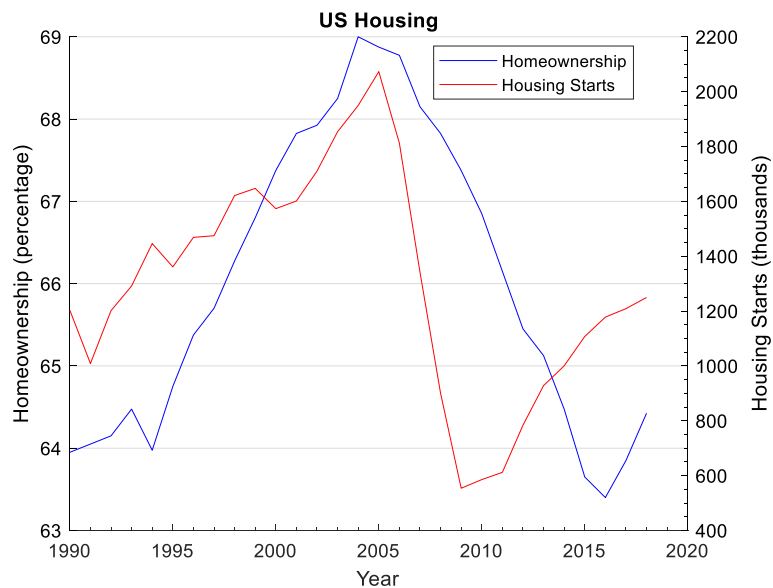

Figure 14. US Gross Output Graph. Trends in US output for differing forestry sectors.

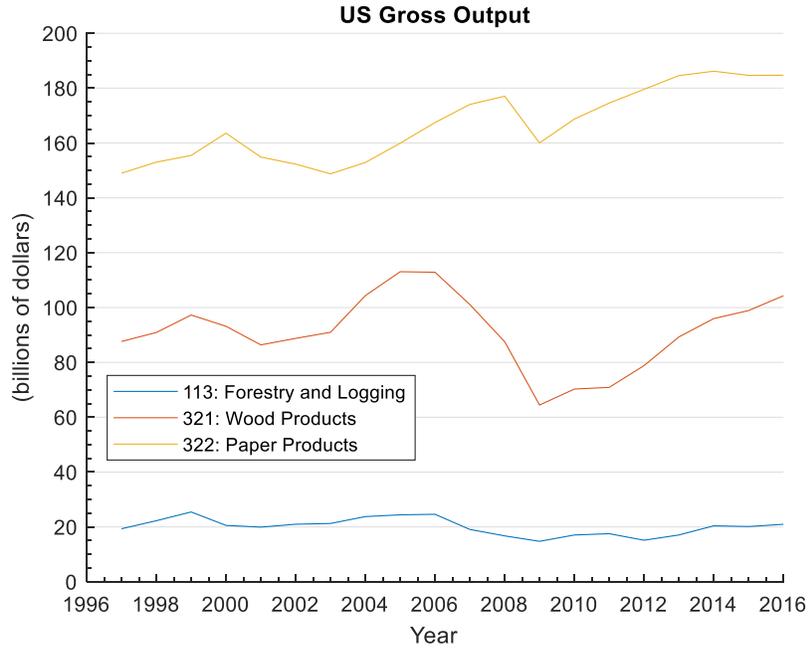

Figure 15. Average Maine Employment Trends in Wood Related Industries Graph. This graph shows employment trends in various traditional wood related industries.

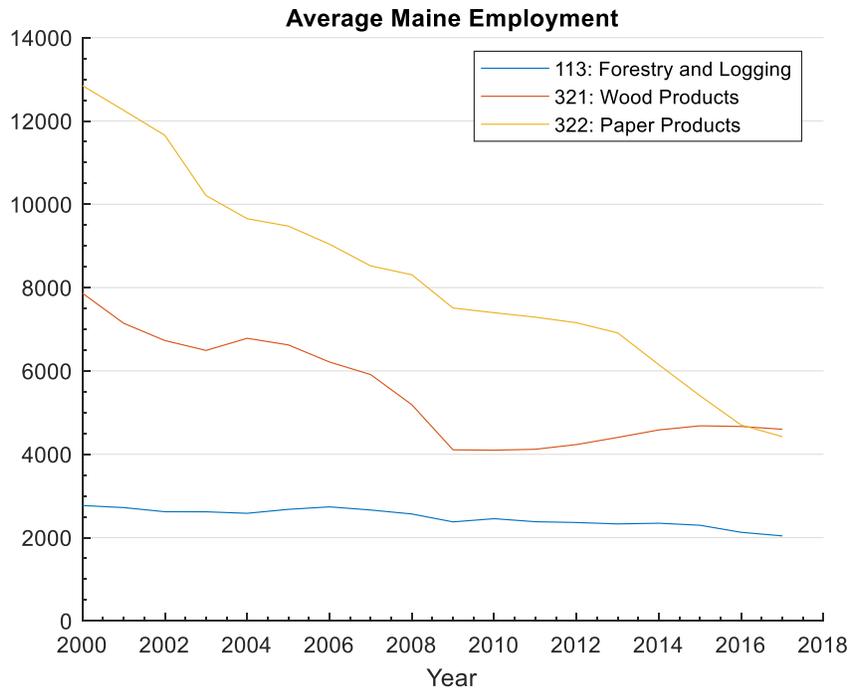

Output in both the forestry and logging industry and forest products industry have decreased during the Great Recession, but have improved since then.US housing starts and US home ownership have significantly dropped since 2006 (and were further worsened by the Great Recession), and thousands of jobs in Maine have been lost (Woodall et al., 2011). Haim et al. (2005) found that 67% of all US wood products were used to construct housing, repair and model, as well as for nonresidential buildings. Baek (2012) found that US lumber price and housing starts are actually more important than the exchange rate between Canada and the US for softwood lumber trade. Short-term fluctuations in US construction markets determine import demand for lumber from Canada, which impacts Canadian forest products (Jennings et al., 1991). Parajuli and Chang (2015) argue public policies must have long term effects since short run price changes (such as temporary tax breaks) won't have an effect in the short run. It is important to note that uncertainty in timber prices influence timber production strategies and periodic dividends paid to timberland shareholders (Mei et al., 2010).

Wood energy in the US has been affected in different ways compared to what we have discussed in this section. Government policies and competing energy prices have affected US wood energy, but the recession did not have an effect (Woodall et al., 2011). With the increased price of competing energy substitutes (primarily electricity) and incentive programs such as the Residential Energy Efficiency Tax Credit, wood energy consumption increased by 29% from 2002 to 2008 (Song et al., 2012; DSIRE, 2011). Aguilar et al. (2011) found other programs also have significant effects on the increase of biomass for energy production which includes the federal renewable energy production tax credit, federal business energy investment tax credits, and Clean Renewable Energy Bonds.

**The Future**

Historically the forest products industry has faced critically important impacts of changes in tariffs, other trade restrictions, changes in the number of operating mills, and changes to demand for forest products (which in turn impacts the forestry and logging industry). Despite historic issues with both industries, harvest volume has been increasing for decades since forest landowners of the north have been properly managing their forests (USDA, 2011). This steady increase in supply could potentially change demand and price of forest products. For these mills that shut down, there have been efforts to repurpose the mills to restore local business for the local economy (primarily in rural areas). Grebner et al. (2009) introduced the field of research that studies alternative uses of low-grade wood, such as production of biofuels and chemical coproducts. Biorefineries would help diversify the industry and broaden the type of end-product uses (compared to pulp, paper or solid wood) (Crandall et al., 2017). Mills aren't just being repurposed for wood energy though, the mills in Old Town and Rumford Maine are currently reopening and being repurposed to softwood by a Chinese firm that can operate at significantly fewer operating costs (compared to the prior firm) (Pendharkar, 2019). In contrast, Augusta transformed their mill into a park to produce amenity instead of tax revenues, this way the negative image of a shutdown didn't lower the value of the area (Irland, 2017).

The Maine Future Forest Economy Project (2005) provided a deeper analysis on what should be done to improve the future of both the forestry and logging industry as well as the forest products industry in Maine beyond just repurposing mills. This report provides a very detailed plan from various industry professionals to improve the industry and connected communities. They argue Maine should encourage investment and improve the business climate via connections and development. This business networking would likely be mutually beneficial in sharing information and learning about emerging opportunities from one another. More specifically, Maine should promote capital and technology investments; this can be achieved by lowering capital taxes, promoting research, development and commercialization of new technologies. Maine should also advocate for diversification of forest products, especially new engineered or bio-products that could provide a competitive advantage due to intellectual property protections. Diversification and marketing campaigns would likely help distinguish Maine products in the global marketplace. A good start would be to highlight Maine's strong spruce-fir resource and strong supply growth. Collaboration would also likely help ensure policy stability and general stability; the University of Maine and Maine state government are great mutually beneficial collaborators that could help the forest products industry in policy, research and development, and community outreach.

  Another important consideration is current and potential US trade partners for lumber imports and exports. Compared with the top 10 countries that the US imports softwood lumber from, Canada makes up 81% of softwood lumber imports for the US (a value of $2.5 billion) (National Association of Home Builders, 2017). Other countries, comprising of 19% of US softwood lumber imports (a value of $600.5 million), include Brazil, Chile, China, New Zealand, Sweden, Mexico, Germany, Romania, and Argentina. What if the US made trade connections with other countries in Europe? Literature shows that parts of Europe are heavy exporters and importers of lumber. Finland and Sweden are common exporters, while Germany and the UK are common importers (Hetemaki et al., 2004). If the US was able to set up regular trade with Germany or the UK, this could help raise demand for forest products which would in turn increase output. Also, since these are all countries in the EU (excluding the UK), this could lead to beneficial trade deals with other countries in the EU as well. An increase in forest product demand would likely strongly benefit suppliers and the local economies of those suppliers.

## Methodology

We analyze the impacts of exchange rate shocks by using a variety of macroeconomic data. Table 1 provides descriptions and sources for each variable and Table 2 provides summary statistics of the listed variables. Figure 4 provides graphs of these variables over time. These variables have reasonable values, range, standard deviation, and no missing values. Quarterly output was proxied from quarterly US RGDP and annual Maine RGDP data. Price is measured as a producer price index for wood products (NAICS 321), and exchange rate is Canadian to US. All other variables are data for Maine's forestry and logging industry which is measured as NAICS 113.

Table 1. Variable List. Includes variable names, descriptions, and sources used to model Maine's forestry and logging industry.

| Variable | Description | Source |
|---|---|---|
| Employment | Employment in Maine for all NAICS 113 (in thousands) | BLS |
| Number of Firms | Employment in Maine for private NAICS 113 | BLS |
| Exchange Rate | Canadian to US Exchange Rate (not seasonally adjusted) | Board of Governors of the Federal Reserve System (US) |
| Price | Producer Price Index for Lumber and Wood Products in 2017 dollars (NAICS 321, not seasonally adjusted) | BLS |
| Wages | Maine's total wage for private NAICS 113 (in thousands) | BLS |
| Output | Proxy of Maine's RGDP (not seasonally adjusted, in millions) | BEA |

Table 2. Summary Statistics. Includes summary statistics of key variables used to model Maine's forestry and logging industry.

| Variable | Mean | Standard Deviation | Minimum | Maximum | N |
|---|---|---|---|---|---|
| Quarter | n/a | n/a | 2001Q1 | 2018Q1 | 69 |
| Employment (in thousands) | 2.513 | 0.164 | 2.067 | 2.800 | 69 |
| Wages | 22817.350 | 4246.641 | 13490.000 | 28600.000 | 69 |
| Exchange Rate | 1.207 | 0.183 | 0.968 | 1.595 | 69 |
| Number of Firms | 467.652 | 31.011 | 420.000 | 539.000 | 69 |
| Price* | 0.852 | 0.096 | 0.665 | 1.078 | 69 |
| Output* | 13408.380 | 397.173 | 12354.850 | 14258.370 | 69 |
| *The following data is cointegrated | | | | | |

Figure 4. Graphs of all variables we used from their time span of 2001Q1-2018Q1.

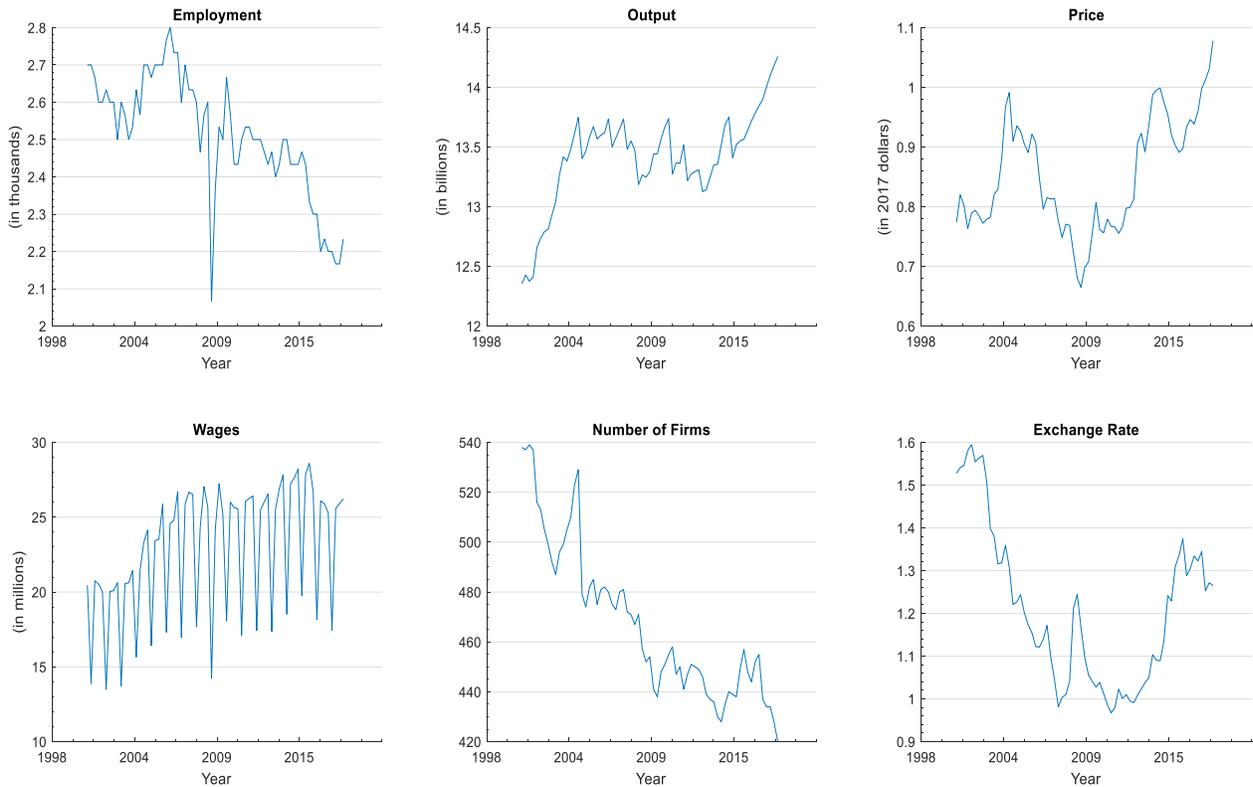

**Model Construction & Justification**

Analyzing the exchange rate shocks and related responses of output and employment require 3 stages of modeling:

(1) Run an in sample VEC model (the same model used in Chapter 2) and then use the initial forecast to shock the exchange rate (see Figure 16).

(2) First difference our variables (including the shocked exchanged rate), and then run an in-sample VAR model holding shocked exchange rate exogenous (which is in and out of sample).

(3) Using the forecasts from Stage 2, run a VAR model (which is both in and out of sample) for all variables holding shocked exchange rate endogenous. Then generate IRFs for the final analysis and interpretation.

Figure 16. Exchange Rate Graph. Graphs the actual exchange rate, the 1$^{st}$ stage forecast, and the shocked exchange rate.

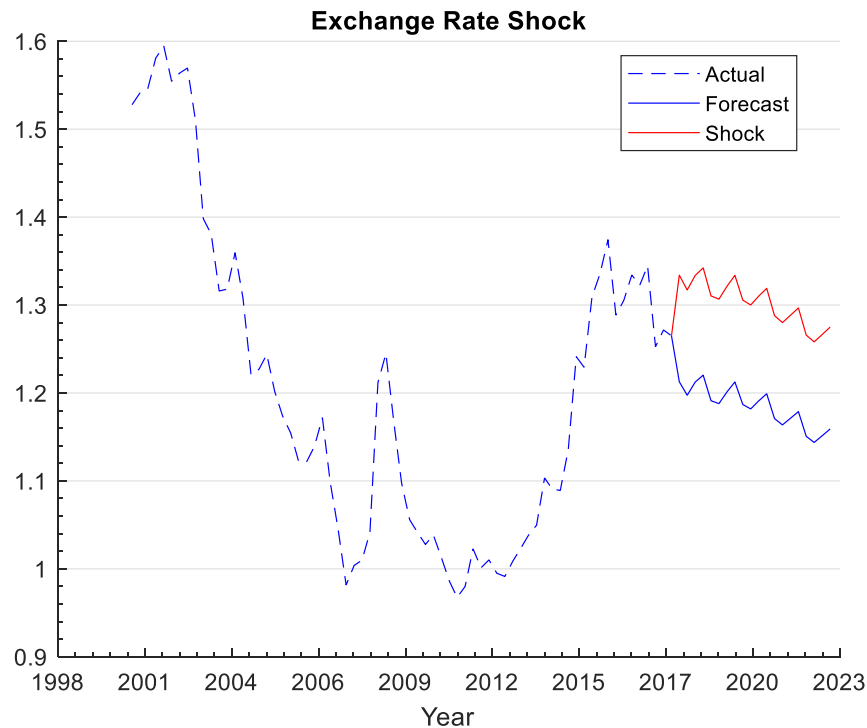

Our VEC model is used instead of a VAR to handle better handle cointegration, and the VECs vector modeling allows us to measure bi-directional and cross directional relationships. This means we can see how everything impacts each other while accounting for everything, which is all accounted for simultaneously. We utilized the following VEC model:

$$\Delta X_t = \sum_{i=1}^{k-1} \Gamma_i \Delta X_{t-i} + \Pi X_{t-1} + \mu + \phi D_t + \epsilon_t$$

$$X_t = [output_t, price_t, employment_t, wage_t, exchangeRate_t, numberFirms_t]'$$

$$t = 2001Q1 - 2018Q1$$

where $X_t$ is a vector of our variables, $k$ is the lag length, $\mu$ is a vector of constant terms, $\phi D_t$ is a vector that measures the number of cointegrating equations (the rank), $\Pi \Delta X_{t-i}$ and $\Gamma X_{t-1}$ are vectors that represent the lag structure, $\Pi$ are long run coefficients, and $\Gamma$ are short run coefficients. The $\Delta X_t$ estimates measure the impact of everything used to analyze the impact of two variables in an IRF, and $\mu$ is used for the Johansen restriction. Our 3-stage modeling process also uses a VAR model:

$$X_t = C + AX_{t-1} + \cdots + AX_{t-k} + \epsilon_t$$

where $X_t$ is the same vector of variables, $k$ is the same number of lags, $C$ is a vector of constants, $A$ is a vector of parameters, and $\epsilon_t$ is a vector of error terms.

VEC and VAR models are both vector models where every variable is at one point the dependent variable with all other variables as independent variables. The primary difference between them are that VECs handle cointegrated data while VARs handle stationary data. The shock to exchange rate was generated by multiplying the VEC models forecasted out of sample exchange rate by 115% (to model a 15% increase starting in 2018Q1). Exchange rate was shocked by a 15% increase to model a 15% increase in tariffs, which has been a regular amount seen historically between the US and Canada (Zhang & Parajuli, 2016; National Association of Home Builders, 2017). This amount can be adjusted in further studies depending on current events, but our 15% was simply chosen due to historic trends. We then first differenced our variables to make them stationary (so we could use them in VAR models for the $2^{nd}$ and $3^{rd}$ stage). This was done so we could hold the in-sample exchange rate and out of sample exchange rate shock exogenous, while we ran the model with all the other variables in the sample. Lastly, we generated IRFs of the $3^{rd}$ stage VAR model that utilized in sample values and the out of sample forecasts from the $2^{nd}$ stage VAR model. These IRFs allow us to analyze the response of employment and output to changes in the exchange rate shock (known as impulses).

**Limitations**

There were issues with data availability. Maine's available data was primarily in annual terms, but there were not enough annual data observations to perform forecasts. We had to proxy Maine's output due to data limitations that has been faced in other literature as well (Daigneault et al, 2016). It is also important to note that the resulting IRFs are calculated from first differenced data (including the first differenced exchange rate shock), so our interpretation will be limited to sign and the direction of the relationship. This is because first differencing data shows changes from one period to the next, so the scale by construction is different. Cointegrated data becomes stationary once first differenced, so our 3 stages of modeling are appropriate. Further research when more regional data regarding supply and demand become readily available would be highly beneficial.

<div align="center">

**Results**

</div>

The IRFs estimated by our 3-stage analysis measure how employment and output in Maine respond to a shock of a 15% increase in exchange rate (from a 15% increase in tariffs) starting in 2018Q1 (Figures 17 and 18). We utilized orthogonalized IRFs which measure an impulse as a one standard deviation change to the impulse variable shocked exchange rate. Orthogonalized IRFs cannot give precise values, but through our 3-stage analysis we were able to provide a precise value for the shock to exchange rate. Output responded to this impulse with a sharp decrease in the first quarter, but then gradually smoothed back to its original levels.

Figure 17. IRF for Shocked Exchange Rate on Employment. Orthogonalized IRF of employments response to an impulse of exchange rate shocked by a 15% increase (all variables are first differenced).

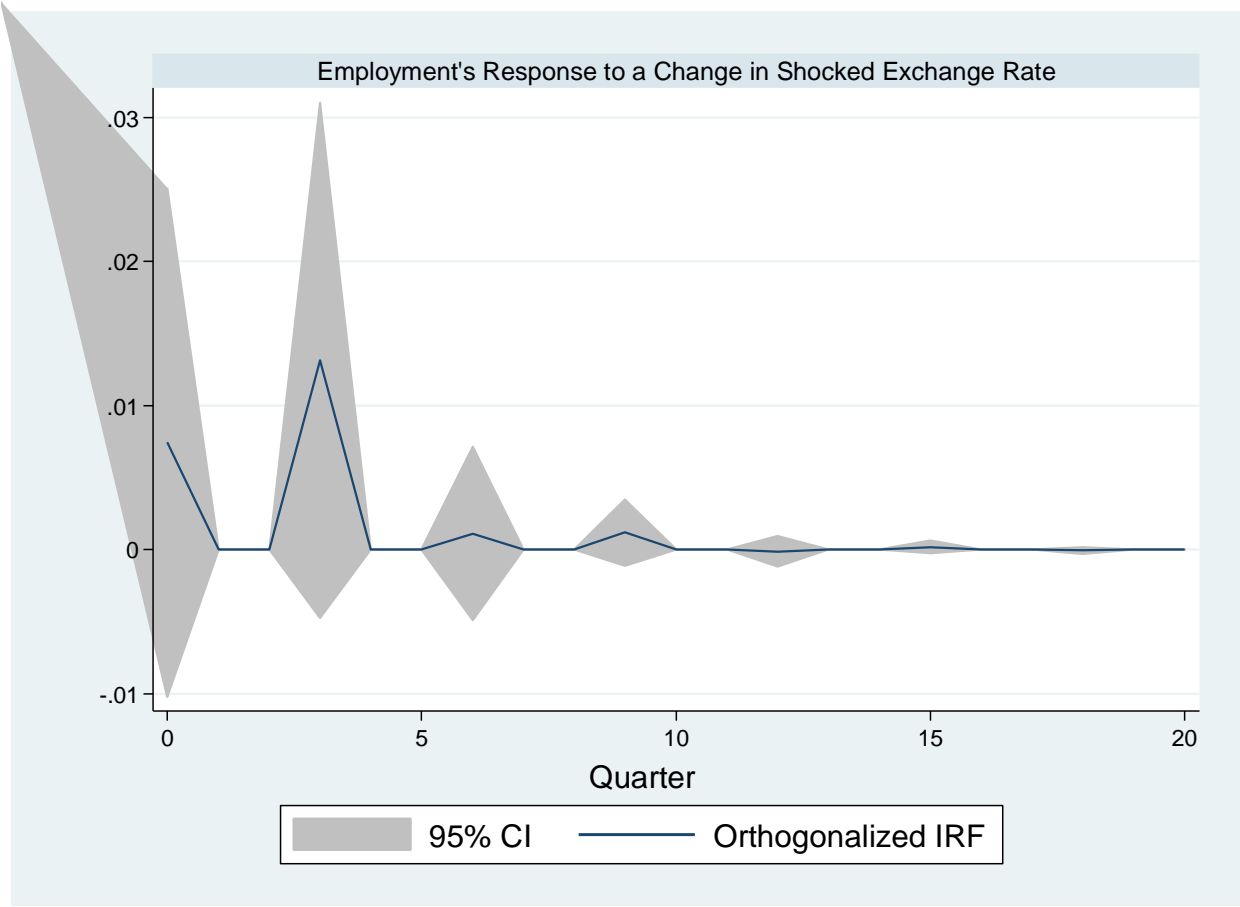

Figure 18. IRF for Shocked Exchange Rate on Output. Orthogonalized IRF of outputs response to an impulse of exchange rate shocked by a 15% increase (all variables are first differenced).

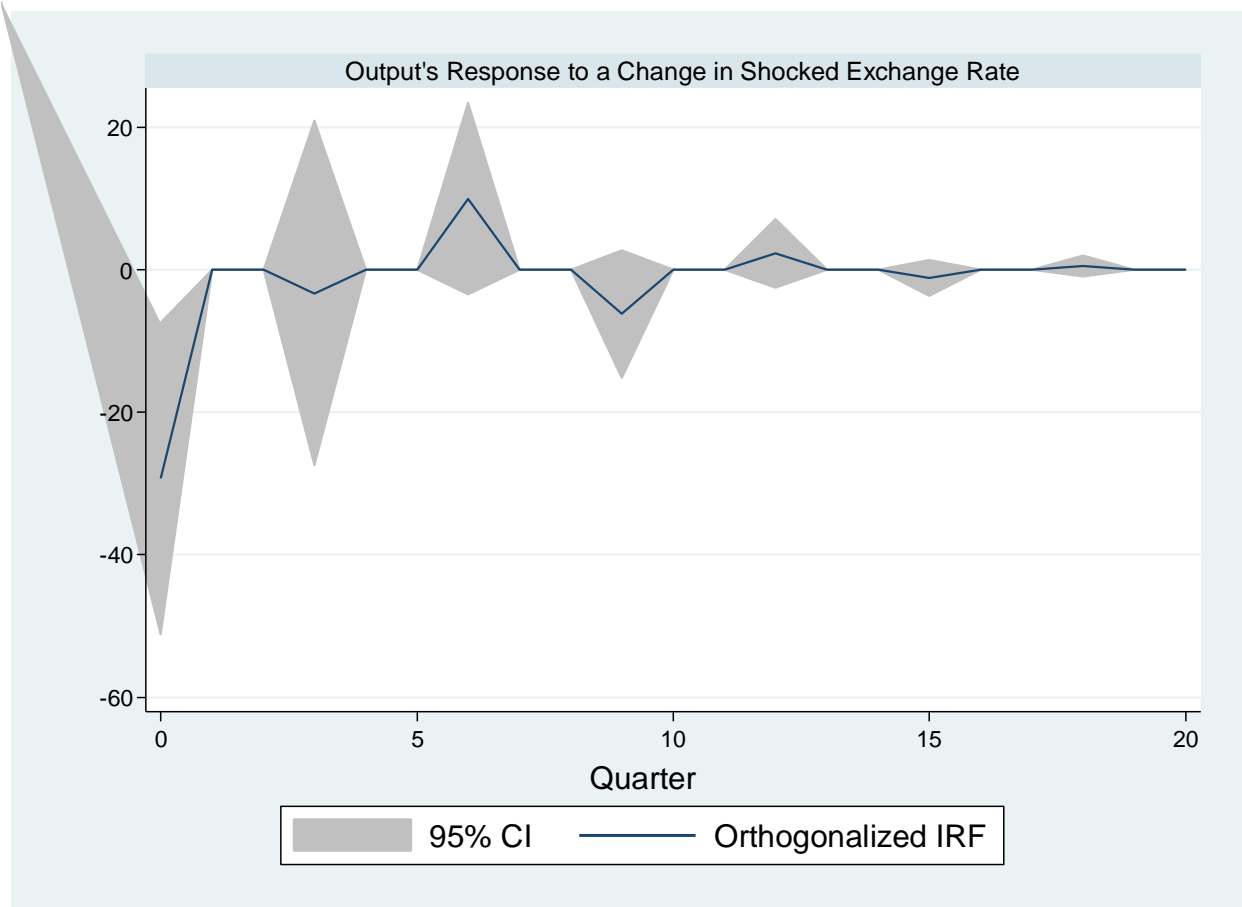

Employment responded to this impulse with oscillating spikes up and down in the first year, but then immediately smoothed back to its original levels.

**Discussion**

The relationships between exchange rate shocks with employment and output in Maine have important implications for Maine's forestry and logging industry. Tariffs that increase the exchange rate by 15% would sharply decrease output in the first quarter, which means that tariffs would restrict output. Restricted output could lower demand, which could accordingly lower employment and the number of firms if they cannot properly adjust to this shock. Output was shown to stabilize after one quarter, so local employment and the number of firms have a good chance to adjust to this shock. This could be due to lead time on orders with Canada. Essentially new orders would be affected by this restriction, but existing orders would not be affected so the existing orders could maintain the industry through this temporary restriction. Global currency markets are also likely playing into this result, but our model does not capture this element (which would be a good element for further study).

We also saw that if the exchange rate increased by 15%, employment would oscillate up and down throughout the first year. These oscillations are never negative; employment temporarily increases and then goes back to its original levels (before spiking again). This is likely due to the short run adjustment process of the model, which simply means it takes time for all variables to reach a new equilibrium after a shock. If Canada stockpiled their orders, then this could be attributed to firms hiring more employees to complete the new increases in orders.

In summary, these IRFs imply that in Maine output could be restricted and employment could face increased volatility. Tariffs essentially make trading goods more expensive and raise uncertainty; these effects could be very harmful to markets, businesses, and related communities (Sun & Zhang, 2003; York, 2018). Uncertainty and volatility are hard to manage in any industry and could negatively affect future investment. There should be serious concern about tariffs increasing in any industry, including both Maine's forestry and logging industry as well as their forest products industry.

## Conclusion

An important part of Maine's economy is the forestry and logging industry. Mill shutdowns and changes in forest products demand have negatively affected local towns (especially rural towns) across the state (as well as the closely related forestry and logging industry). This is because many depend on mills for employment of mill workers, harvesters, loggers, and support staff (Weeks, 1990). These towns aren't just negatively affected by decreases in related employment, but also general business activity and even the tax base. There is an important trade dynamic between the US and Canada for which exchange rate has a critical role (Hietala et al., 2013). When the exchange rate appreciates, US firms have a harder time exporting, so exchange rate can be used as a general indicator (Irland, 2017). Tariffs and other trade restrictions have a significant negative impact on US imports from Canada, the dominant importer for the US (Nagubadi & Zhang, 2013). Historic Canadian lumber exports with the US show tariffs and other trade restrictions have significant effects on the number of exports (Zhang & Parajuli, 2016).

  To analyze the effect of tariff increases, we shocked exchange rate by 15% starting in 2018Q1 and then analyzed the responses of output and employment in Maine. It is important to analyze, since this can show policy makers the implications of changing tariffs and trade restrictions. We completed 3 stages of modeling that made use of a combination of VEC and VAR models to generate IRFs based on in and out of sample variable values. We used the same quarterly data from Chapter 2 which spanned from 2001Q1 to 2018Q1 and forecasts up until 2023Q1. Most of Maine's data is primarily annual (which doesn't provide enough observations), but our available quarterly data provided more observations. This research provides a significant contribution to research in both the forestry and logging industry as well as the forest products industry since it provides the best data and methodology currently available. The IRFs imply that if exchange rate increased 15% (due to tariffs increasing) then output would be restricted, and employment could face increased volatility. Increasing tariffs in any industry could be very harmful due to restricted output and uncertainty, both markets, businesses, and related communities would likely be negatively affected.

      This shock analysis is especially important for both Maine's forestry and logging industry and the forest products industry. Many of Maine's local economies active in both the forestry and logging industry and forest products industry have struggled during the past 20 years with 64% of their mills shutting down. This has been especially difficult for rural communities with tax bases dependent on these mills. Mill owners and policymakers (especially in Maine) should take care if they consider changing tariffs, they have important effects on employment and output in the overall economy and especially local economies. This analysis could be used for both the forestry and logging industry and the forest products industry in different states or regions across the country. We especially encourage further studies in other areas that have found to be important in the literature such as the Southeast and Northwest of the US, as well as other countries including Canada, Sweden, Germany, and Finland (Hetemaki et al, 2004; Malaty et al, 2006; Zhang & Parajuli, 2016; National Association of Home Builders, 2017). The EU is another important area of study due to their efficiencies of trade within which could also benefit any foreign traders. This analysis has set up a framework for modeling shocks to macroeconomic forecasts, providing insight in the impacts of recent US government tariffs and trade restrictions at a regional level for a specific industry.

# CHAPTER 4: CONCLUSIONS

My thesis research answered two questions: (1) How will key macroeconomic and related variables for Maine's forestry and logging industry change in the future? (2) How will shocks to exchange rates from an increase in tariffs (and other trade restrictions) affect employment and output in Maine's forestry and logging industry? Our forecasting results for Maine imply local communities could be worse off due to the number of firms and employment decreasing (even though the industry will likely remain stable due to level prices and a slight increase in output). In response to a 15% increase in exchange rate (due to increasing tariffs and other trade restrictions), our results show output in Maine would restrict and Maine employment would face increased volatility. These findings suggest policymakers and mill owners should carefully consider if they should change tariffs (since this could lead to uncertainty and increased costs of business operations). In answering these questions, this research has set a framework for a macroeconomic approach to analyzing both the forestry and logging industry and forest products industry by forecasting and shock analysis (with according responses).

We developed an improved process for forecasting macroeconomic and related variables in Maine's forestry and logging industry. Our data spanned from 2001Q1 to 2018Q1, and we forecasted up until 2023Q1 with a VEC model. This research properly accounts for the cointegrated nature of US lumber data and significantly contributes to research for both the forestry and logging industry and forest products industry. Our forecasts show that although price and output are forecasted to stay stable in Maine, employment and the number of firms are forecasted to decrease. So, even though the forestry and logging industry is forecasted to be stable in Maine, local communities will be harmed (especially rural communities with tax bases dependent on mills and related employment). Even though this paper only analyzed the industry in Maine, this type of analysis can also be used for different states or regions across the country. This analysis provides important information about the general economy and more specific local economies that should be carefully considered by policymakers and mill owners.

Making use of the data and model built in Chapter 2, Chapter 3 analyzes the effect of tariff increases. To do this, we shocked exchange rate by 15% starting in 2018Q1 and then analyzed the responses of output and employment in Maine. This was done through 3 stages of modeling that used both VEC and VAR models to generate IRFs based on in and out of sample variable values. We made use of both VEC and VAR models to exogenously handle the initial in sample exchange rate values and out sample exchange rate shock (while all other in sample variables were held endogenous). The IRFs of exchange rate increasing 15% (due to tariffs) imply employment volatility and output restrictions in Maine. This could negatively affect businesses and markets in Maine due to increased uncertainty and increased costs of business operations, which could negatively impact related communities, businesses and markets. Mill owners and policymakers (especially in Maine) should be careful if they consider changing tariffs and other trade restrictions, because these have important effects on employment and output in the overall economy and especially local economies.

In summary, this research of Maine's forestry and logging industry forecasted key macroeconomic and related variables, and then analyzed how employment and output in Maine would respond to increases in tariffs. This provides critical insight to the direction of the economy of this industry, the related economy, and how changes in tariffs can impact all of these. Policymakers and mill owners should carefully consider our forecasts and analyses as they adjust to a perpetually changing industry and tariff appreciations recently implemented by the US government. This research sets up a framework of a macroeconomic approach to analyzing both the forestry and logging industry and forest products industry. We highly encourage future research to analyze other regions of the US (such as the southeast and northwest) as well as other countries found to be important in the literature such as Canada, Sweden, Germany, and Finland (Hetemaki et al, 2004; Malaty et al, 2006; Zhang & Parajuli, 2016; National Association of Home Builders, 2017). Countries in the EU is a particularly important to analyze due to their efficiencies of trade between each other, which could also benefit any traders outside of the EU.

# BIOGRAPHY OF THE AUTHOR

On March 25th, 1994, Jonathan E. Gendron was born in Lanoka Harbor, NJ. He graduated from Wilde Lake High School in Columbia, MD in 2012. He then attended Towson University in Towson, MD in 2012 and then in 2016 earned a Bachelors of Science in Economics with honors *cum laude* and minored in Computer Information Systems. He then enrolled in the School of Economics at the University in Maine in 2017. His experience involves economic impact analyses, forecasting and applied econometrics. He is contributing to research for Maine's forestry and logging industry by forecasting key macroeconomic and related variables and performing shock analyses. Jonathan is a candidate for the Master of Science degree in Economics from the University of Maine in May 2019.